\documentclass[aps,prd,preprint,floats,epsf,superscriptaddress]{revtex4}
\usepackage{amsfonts}
\usepackage{amsmath}

\begin{document}

\title{Gauge-Higgs unification models in six dimensions with $S^2/Z_2$ extra space and GUT gauge symmetry}
\date{\today}

\author{Cheng-Wei Chiang}
\email[e-mail: ]{chengwei@ncu.edu.tw}
\affiliation{Department of Physics and Center for Mathematics and Theoretical Physics, National Central University, Chungli, Taiwan 320, R.O.C.}
\affiliation{Institute of Physics, Academia Sinica, Taipei, Taiwan 115, R.O.C.}
\author{Takaaki Nomura}
\email[e-mail: ]{nomura@ncu.edu.tw}
\affiliation{Department of Physics and Center for Mathematics and Theoretical Physics, National Central University, Chungli, Taiwan 320, R.O.C.}  
\author{Joe Sato}
\email[e-mail: ]{joe@phy.saitama-u.ac.jp}
\affiliation{Department of Physics,
  Saitama University,
  Shimo-Okubo,
  Sakura-ku,
  Saitama 355-8570, Japan} 

\date{\today}

\begin{abstract}
In this article, we review gauge-Higgs unification models based on gauge theories defined on six-dimensional spacetime with $S^2/Z_2$ topology in the extra spatial dimensions.
On the extra $S^2/Z_2$ space, non-trivial boundary conditions are imposed.
This review considers two scenarios for constructing a four-dimensional theory from the six-dimensional model. 
One scheme utilizes the SO(12) gauge symmetry with a special symmetry condition imposed on the gauge field, whereas the other employs the E$_6$ gauge symmetry without requiring the additional symmetry condition.
Both models lead to a Standard Model-like gauge theory with the SU(3) $\times$ SU(2)$_L$ $\times$ U(1)$_Y$($\times$ U(1)$^2$) symmetry and SM fermions in four dimensions.
The Higgs sector of the model is also analyzed.  The electroweak symmetry breaking can be realized, and the weak gauge boson and Higgs boson masses are obtained.
\end{abstract}

\maketitle


%
%
\section{Introduction}
The Higgs sector of the Standard Model (SM) plays an essential role in the spontaneous symmetry breaking (SSB) from the SU(3)$_C$ $\times$ SU(2)$_L$ $\times$ U(1)$_Y$ gauge group down to SU(3)$_C$ $\times$ U(1)$_{EM}$, thereby giving masses to the SM elementary particles. 
However, the SM does not address the most fundamental nature of the Higgs sector, such as the mass and self-coupling constant of the Higgs boson.
Therefore, the Higgs sector is not only the last territory in the SM to be discovered, but will also provide key clues to new physics at higher energy scales.


Gauge-Higgs unification is one of many attractive approaches to physics beyond the SM in this regard~\cite{%
  Manton:1979kb,%
  Fairlie:1979at,%
  Fairlie:1979zy%
} (for recent approaches, 
 see Refs.~\cite{%
  Hall:2001zb,%
  Burdman:2002se,%
  Gogoladze:2003ci,%
  Scrucca:2003ut,%
  Haba:2004qf,%
  Biggio:2004kr,%
  Maru:2004,%
  Haba:2005kc,%
  Hosotani:2006qp,%
  Sakamoto:2006wf,%
  Maru:2006,%
  Hosotani:2007qw,%
  Sakamura:2007qz,%
  Medina:2007hz,%
  maru2007,%
  Adachi:2007tj,%
  Gogoladze:2007ey%
}). 
In this approach, the Higgs particles originate from the extra-dimensional components of the gauge field defined on spacetime with the number of dimensions greater than four.
In other words, the Higgs sector is embraced into the gauge interactions in the higher-dimensional model, and many fundamental properties of Higgs boson are dictated by the gauge interactions.
%


In our recent studies, we have shown interesting properties of one type of gauge-Higgs unification models based on grand unified gauge theories defined on six-dimensional (6D) spacetime, with the extra-dimensional space having the topological structure of two-sphere orbifold $S^2/Z_2$~\cite{Nomura:2008sx, Chiang:2010hy}.

In the usual coset space dimensional reduction (CSDR) approach~\cite{Manton:1979kb,Forgacs:1979zs,Kapetanakis:1992hf,Chatzistavrakidis:2007by,Zoupanos08}, one imposes on the gauge fields the symmetry condition which identifies the gauge transformation as the isometry transformation of $S^2$ due to its coset space structure $S^2$=SU(2)/U(1).
The dimensional reduction is explicitly carried out by applying the solution of the symmetry condition.  A background gauge field is introduced as part of the solution~\cite{Manton:1979kb}.  Such a background gauge field is also necessary for obtaining chiral fermions in four dimensional (4D) spacetime, even without the symmetry condition.
After the dimensional reduction, no Kaluza-Klein (KK) mode appears because of the imposed symmetry condition.
%
%
The symmetry condition also restricts the gauge symmetry and the scalar contents originated from the extra gauge field components in the 4D spacetime.
Moreover, a suitable potential for the scalar sector can be obtained to induce SSB at tree level.

In this article, we consider two scenarios for constructing the 4D theory from a 6D model: 
one utilizing the symmetry condition for the gauge field, whereas the other without it.
In the first scenario, however, we do not impose the condition on the fermions as used in other CSDR models.
We then have massive KK modes for fermions but not the gauge and scalar fields in 4D.  We can thus obtain a dark matter candidate under assumed KK parity.
In the case without the symmetry condition, we find that the background gauge field restricts the gauge symmetry and massless particle contents in 4D.
Also, there are KK modes for each field, with the mass spectrum determined according to the model.
Generally, massless modes do not appear in the KK mass spectrum because of the positive curvature of the $S^2$ space~\cite{A.A.Abrikosov}.
With the help of the background gauge field, however, we obtain massless KK modes for the gauge bosons and fermions.

Generally, the gauge symmetry of a grand unified theory (GUT) tends to remain in 4D in these dimensional reduction approaches  
~\cite{Kapetanakis:1992hf,Chapline:1982wy,10dim-Model:K12,10dim-Model:D14,10dim-Model:B,CSDR14D}.
Also, it is difficult to obtain the appropriate Higgs potential to break the GUT gauge symmetry because of the gauge group structure.
A GUT gauge symmetry can be broken to the SM gauge symmetry by the non-trivial boundary conditions (for cases with orbifold extra space, see, for example,~\cite{%
  Hall:2001zb,%
  Burdman:2002se,%
  Gogoladze:2003ci,%
  Scrucca:2003ut,%
  Haba:2004qf,%
  Haba:2005kc,%
  Hosotani:2006qp,%
  Sakamura:2007qz,%
  Medina:2007hz,%
  maru2007,%
  Kawamura1,%
  Kawamura2}).
Therefore, to solve the above-mentioned problems, we impose on the fields of the 6D model a set of non-trivial boundary conditions on the $S^2/Z_2$ space.  Therefore, the gauge symmetry, scalar contents and massless fermions are determined by these boundary conditions and the background gauge field.  We find that in both scenarios, with or without the symmetry condition for the gauge field, the electroweak symmetry breaking (EWSB) can be realized, 
and the Higgs boson mass is predicted by analyzing the Higgs potential in the respective model.


This paper is organized as follows.
In Sec.~\ref{CSDR}, we review the two scenarios for constructing the 4D theory from gauge theories on 6D spacetime whose extra space has the $S^2/Z_2$ topology with a set of 
non-trivial boundary conditions.
In Sec.~\ref{models}, we show the models based on SO(12) and E$_6$ gauge symmetries, with the former imposed by the symmetry condition on the gauge field and the latter without.
We summarize our results in Sec.~\ref{summary}.



\section{The 6D Gauge-Higgs Unification model construction scheme with extra-space $S^2/Z_2$}
\label{CSDR}

In this section, we recapitulate the scheme for constructing a 4D theory from a gauge theory on 
6D spacetime which has extra-space as two-sphere orbifold $S^2/Z_2$. 
We consider two cases; one case uses the symmetry condition and the other case does not use symmetry condtion.
We apply non-trivial boundary condition for both cases.


\subsection{A Gauge theory on 6D spacetime with $S_2/Z_2$ extra-space}

\subsubsection{\bf The 6D spacetime with $S_2/Z_2$ extra-space}
We begin with properties of a 6D spacetime $M^6$.
The spacetime $M^6$ is assumed to be a direct product of the 4D 
Minkowski spacetime $M^4$ and two-sphere orbifold $S^2/Z_2$ such that $M^6=M^4 \times S^2/Z_2$.
The two-sphere $S^2$ is a unique two-dimensional coset space, and can be written 
as $S^2 = \mathrm{SU}(2)_I/\mathrm{U}(1)_I$, where U(1)$_I$ is the subgroup of SU(2)$_I$.
This coset space structure of $S^2$ requires 
that $S^2$ has the isometry group SU(2)$_I$, and that the group U(1)$_I$ is embedded into 
the group SO(2) which is a subgroup of the Lorentz group SO(1,5).
We denote the coordinate of $M^6$ by $X^M=(x^{\mu},y^{\theta}=\theta,y^{\phi}=\phi)$,
where $x^{\mu}$ and $\{ \theta,\phi \}$ are $M^4$ coordinates and $S^2$ spherical coordinates,
 respectively.
The spacetime index $M$ runs over $\mu$ $\in$ $\{ 0,1,2,3 \}$ and $\alpha$ $\in$ $\{ \theta,\phi \}$.
The orbifold $S^2/Z_2$ is defined by the identification of $(\theta,\phi)$ and $(\pi - \theta,-\phi)$~\cite{Maru:2009wu}.  The two fixed points are $(\pi/2,0)$ and $(\pi/2,\pi)$.
The metric of $M^6$, denoted by $g_{MN}$, can be written as
\begin{equation}
g_{MN} = \begin{pmatrix} \eta_{\mu \nu} & 0 \\ 0 & -g_{\alpha \beta} \end{pmatrix}, 
\end{equation}
where $\eta_{\mu \nu}= diag(1,-1,-1,-1)$ and $g_{\alpha \beta}= diag(1, \sin^{-2} \theta)$ 
are metric of $M^4$ and $S^2$ respectively.
Notice that we omit the radius $R$ of $S^2$ in this discussion. 
We define the vielbein $e^{M}_{A}$ 
that connects the metric of $M^6$ and that of 
the tangent space of $M^6$, denoted by $h_{AB}$, as $g_{MN}=e_M^{A} e_N^B h_{AB}$. 
Here $A=(\mu,a)$, where $a$ $\in$ $\{ 4,5 \}$, is the index for the coordinates of tangent space of $M^6$. 
The explicit form of the vielbeins are summarized as 
\begin{equation}
e^1_{\theta} = 1, \quad
e^2_{\phi} = \sin \theta, \quad
e^1_{\phi} = e^2_{\theta} = 0. 
\end{equation}
Also the non-zero components of spin connection are 
\begin{equation}
R^{12}_{\phi} = - R^{21}_{\phi} = -\cos \theta.
\end{equation}


\subsubsection{\bf A Lagrangian on 6D spacetime with $S_2/Z_2$ extra-space}
We then show the general structure of a gauge theory on $M^6$.
We introduce a gauge field $A_{M}(x,y)=(A_{\mu}(x,y),A_{\alpha}(x,y))$, which belongs to 
the adjoint representation of the gauge group $G$, and fermions $\psi(x,y)$, which lies in 
a representation $F$ of $G$.
The action of this theory is given by 
\begin{equation}
\label{6Daction}
S = \int dx^4 \sin \theta d \theta d \phi \bigl(
\bar{\Psi} i \Gamma^{\mu} D_{\mu} \Psi + \bar{\Psi} i \Gamma^{a} e^{\alpha}_{a} D_{\alpha} \Psi 
- \frac{1}{4 g^2} g^{MN} g^{KL} Tr[F_{MK} F_{NL}] \bigr) ,
\end{equation}
where $F_{MN}= \partial_M A_N(X) -\partial_N A_M(X) -[A_M(X),A_N(X)]$ is the field strength, 
$D_M$ is the covariant derivative including spin connection, and $\Gamma_A$ represents 
the 6-dimensional 
Clifford algebra.
Here $D_M$ and $\Gamma_A$ can be written explicitly as,
\begin{align}
& D_{\mu} = \partial_{\mu} - A_{\mu}, \quad
D_{\theta} = \partial_{\theta} - A_{\theta}, \quad
\label{Dphi}
D_{\phi} = \partial_{\phi} -i \frac{\Sigma_3}{2} \cos \theta -A_{\phi}, \\
& \Gamma_{\mu} = \gamma_{\mu} \otimes \mathbf{I}_2, \quad
\Gamma_4 = \gamma_{5} \otimes \sigma_1, \quad
\Gamma_5 = \gamma_{5} \otimes \sigma_2, 
\end{align}
where $ \{ \gamma_{\mu}, \gamma_{5} \} $ are the 4-dimensional Dirac matrices, 
$\sigma_i(i=1,2,3)$ are Pauli matrices, $\mathbf{I}_d$ is $d \times d$ identity, 
and 
$\Sigma_3$ is defined as $\Sigma_3=\mathbf{I}_4 \otimes \sigma_3$.
The covariant derivative $D_{\phi}$ has the spin connection term $i \frac{\Sigma_3}{2} \cos \theta$ which is needed for space with non-zero curvature like $S^2$ and 
applied for only fermions.
In 6D spacetime, we can define chirality of fermions and the projection operators are defined as 
\begin{eqnarray}
\Gamma_{\pm} = \frac{1 \pm \Gamma_7 }{2}
\end{eqnarray}
where $\Gamma_7 \equiv \gamma_5 \otimes \sigma_3$ is the chiral operator.
We can write chiral fermion on 6D spacetime such as 
\begin{eqnarray}
\Psi_{\pm} = \Gamma_{\pm} \Psi, \qquad
\Gamma_7 \Psi_\pm = \pm \Psi_\pm
\end{eqnarray}
where $\psi$ is a Dirac fermion on 6D spacetime.
The 6D chiral fermion can be also written in terms of 4D chiral fermions $\psi_{L(R)}$ as 
\begin{eqnarray}
\label{chiral1}
\Psi_{+} &=& \begin{pmatrix} \psi_{R} \\ \psi_{L} \end{pmatrix}, \\
\label{chiral2}
\Psi_{-} &=& \begin{pmatrix} \psi_{L} \\ \psi_{R} \end{pmatrix}.
\end{eqnarray} 


\subsubsection{\bf Non-trivial boundary conditions on the two-sphere orbifold \label{sec:bc}}

On the two-sphere orbifold, one can consider parity operations $P: \, (\theta,\phi) \to (\pi-\theta,-\phi)$ and azimuthal translation $T_\phi: \, (\theta,\phi) \to (\theta,\phi+2\pi)$.  Here we note that the periodicity $\phi \to \phi+2\pi$ is not associated with the orbifolding.  We can impose the following two types of boundary conditions on both gauge and fermion fields under the two operations:
\begin{eqnarray}
\label{boundary-condition1}
A_{\mu} (x,\pi-\theta,-\phi) 
&=& P_1 A_{\mu}(x,\theta,\phi) P_1 ~, \
A_{\mu} (x,\pi-\theta,2\pi-\phi) 
= P_2 A_{\mu}(x,\theta,\phi) P_2 ~, \\
\label{boundary-condition2}
A_{\alpha}(x,\pi-\theta,-\phi) 
&=& - P_1 A_{\theta,\phi}(x,\theta,\phi) P_1 ~, \
A_{\alpha}(x,\pi-\theta,2\pi-\phi) 
= - P_2 A_{\alpha}(x,\theta,\phi) P_2 ~, \\
\label{boundary-condition3}
\Psi (x,\pi-\theta,-\phi) 
&=& \pm \gamma_5 P_1 \Psi(x,\theta,\phi) ~, \
\Psi (x,\pi-\theta,2\pi-\phi) = \pm \gamma_5 P_2 \Psi(x,\theta,\phi) ~,
\end{eqnarray}
or
\begin{eqnarray}
\label{boundary-condition4}
A_{\mu} (x,\pi-\theta,-\phi) 
&=& P_1 A_{\mu}(x,\theta,\phi) P_1 ~, \
A_{\mu} (x,\theta,\phi+2\pi) 
= P_2 A_{\mu}(x,\theta,\phi) P_2 ~, \\
\label{boundary-condition5}
A_{\alpha}(x,\pi-\theta,-\phi) 
&=& - P_1 A_{\theta,\phi}(x,\theta,\phi) P_1 ~, \
A_{\alpha}(x,\theta,\phi+2\pi) 
=  P_2 A_{\alpha}(x,\theta,\phi) P_2 ~, \\
\label{boundary-condition6}
\Psi (x,\pi-\theta,-\phi) 
&=& \pm \gamma_5 P_1 \Psi(x,\theta,\phi) ~, \
\Psi (x,\theta,\phi+2\pi) = \pm P_2 \Psi(x,\theta,\phi) ~,
\end{eqnarray}
where former conditions are associated with $P$ operation and combination of $P$ and $T_\phi$ operations, and latter conditions are associated with $P$ and $T_\phi$ operation itself.
These boundary conditions are determined by requiring the invariance of the six -dimensional action under the transformation $(\theta,\phi) \rightarrow (\pi-\theta,-\phi)$ and $\phi \rightarrow \phi+2\pi$.

The projection matrices $P_{1,2}$ act on the gauge group representation space and have eigenvalues $\pm 1$.  They assign different parities for different representation components.  For fermion boundary conditions, the sign in front of $\gamma_5$ can be either $+$ or $-$ since the fermions always appear in bilinear forms in the action.  The 4-dimensional action is then restricted by these parity assignments.



\subsection{The dimensional reduction scheme with symmetry condtion}

Here we review the dimensional reduction scheme which apply symmetry condition of gauge field~\cite{Nomura:2008sx}.

\subsubsection{\bf The symmetry condition}
 
We impose on the gauge field $A_M(X)$   
the symmetry which connects SU(2)$_I$ isometry transformation on $S^2$
and the gauge transformation on the fields in order to carry out dimensional reduction, and 
the non-trivial boundary conditions of $S^2/Z_2$ to restrict 4D theory.
The symmetry requires that the SU(2)$_I$ coordinate transformation should be compensated by
a gauge transformation~\cite{Manton:1979kb,Forgacs:1979zs}.
The symmetry further leads to the following set of the symmetry condition on the fields:
\begin{align}
\label{symm-con-vec4} 
\xi_i^{\beta} \partial_{\beta} A_{\mu} 
&= \partial_{\alpha} W_i + [W_i,A_{\mu}],   \\
\label{symm-con-vecex}
\xi_i^{\beta} \partial_{\beta} A_{\alpha} + \partial_{\alpha} \xi_i^{\beta} A_{\beta} 
&= \partial_{\alpha} W_i + [W_i,A_{\alpha}],    
\end{align}
where $\xi_i^{\alpha}$ is the Killing vectors generating SU(2)$_I$ symmetry and $W_i$ are some fields which 
generate an infitesimal gauge transformation of $G$.
Here index $i = 1,2,3$ corresponds to that of SU(2) generators. 
The explicit forms of $\xi_i^{\alpha}$s for $S^2$ are:
\begin{align}
\xi_1^{\theta} &= \sin \phi ,  \qquad \xi_1^{\phi} = \cot \theta \cos \phi, \nonumber \\
\xi_2^{\theta} &= -\cos \phi ,  \qquad \xi_2^{\phi} = \cot \theta \sin \phi, \nonumber \\
\xi_3^{\theta} &= 0 ,  \qquad \xi_3^{\phi} = -1. 
\end{align} 
The LHSs of Eq~(\ref{symm-con-vec4},\ref{symm-con-vecex}) are infintesimal isometry SU(2)$_I$ transformation and 
the RHSs of those are infintesimal gauge transformation.
%

%
%
%
%
\subsubsection{\bf The dimensional reduction and a Lagrangian in 4D spacetime}
The dimensional reduction of gauge sector is explicitly carried out by applying the solutions of 
the symmetry condition Eq~(\ref{symm-con-vec4},\ref{symm-con-vecex}).
These solutions are given by Manton~\cite{Manton:1979kb} as 
\begin{align}
\label{kaiA}
&A_{\mu} = A_{\mu}(x), \quad
A_{\theta} = -\Phi_1(x), \quad
A_{\phi} = \Phi_2(x) \sin \theta - \Phi_3 \cos \theta, \\ 
\label{solW1}
&W_1 = - \Phi_3 \frac{\cos \phi}{\sin \theta}, \quad
W_2 = - \Phi_3 \frac{\sin \phi}{\sin \theta}, \quad
W_3 = 0,
\end{align}
and satisfy the following constraints:
\begin{align}
\label{kousoku1}
[\Phi_3,A_{\mu}] &= 0, \\ 
\label{kousoku2}
[-i \Phi_3,\Phi_i(x)] &= i \epsilon_{3ij} \Phi_j(x), 
\end{align}
where $\Phi_1(x)$ and $\Phi_2(x)$ are scalar fields, and $-i\Phi_3$ are chosen as generator of U(1)$_I$. 
Note that the $\Phi_3$ term for $A_\phi$ corresponds to the background gauge field~\cite{background}.
Substituting the solutions Eq~(\ref{kaiA}) into $A_M(X)$ in action Eq~(\ref{6Daction}), 
we can easily integrate coordinates $\theta$ and $\phi$ in 
the gauge sector.
We then obtain a four dimensional action as 
\begin{align}
\label{4d-action}
S_{4D}^{(gauge)} = \int d^4x 
\biggl( 
&- \frac{1}{4g^2} Tr[F_{\mu \nu} F^{\mu \nu}(x)] \nonumber \\
&- \frac{1}{2g^2} Tr[D'_{\mu}\Phi_1(x) D'^{\mu} \Phi_1(x)+D'_{\mu}\Phi_2(x) D'^{\mu} \Phi_2(x)] \nonumber \\
&- \frac{1}{2g^2} Tr[(\Phi_3+[\Phi_1(x),\Phi_2(x)])(\Phi_3+[\Phi_1(x),\Phi_2(x)])]  
\biggr), 
\end{align}
where $D'_{\mu} \Phi = \partial_{\mu} -[A_{\mu},\Phi]$.
The fermion sector of 4D action is obtained by expanding fermions 
in normal modes of $S^2/Z_2$ and then integrating $S^2/Z_2$ coordinate in 6D action.
Thus,
 the fermions have massive KK modes which would be a candidate of dark matter.
Generally, the KK modes do not have massless mode because of 
the positive curvature of $S^2$~\cite{A.A.Abrikosov}.
We, however, can show that the fermion components satisfying the following condition have massless mode:
\begin{equation} 
\label{kousoku3}
-i \Phi_3 \psi = \frac{\Sigma_3}{2} \psi.
\end{equation}
Square mass of the KK modes are eigenvalues of square of extra-dimensional Dirac-operator $-i \hat{D}$.
In the $S^2$ case, $-i \hat{D}$ is written as
\begin{align}
\label{dirac}
-i \hat{D} &= -i e^{\alpha a} \Gamma_a D_{\alpha} \nonumber \\ 
&=-i \bigl[ \Sigma_1 (\partial_{\theta} + \frac{\cot \theta}{2} ) 
+ \Sigma_2 (\frac{1}{\sin \theta} \partial_{\phi} + \Phi_3 \cot \theta ) \bigr],
\end{align}
where $\Sigma_i=\mathbf{I}_4 \times \sigma_i$.
Square of $-i \hat{D}$ can be explicitly calculated:
\begin{align}
\label{dirac-square}
(-i \hat{D})^2 = - \bigl[ \frac{1}{\sin \theta} \partial_{\theta} (\sin \theta \partial_{\theta}) 
+ \frac{1}{\sin^2 \theta} \partial_{\phi}^2 +i (2(-i\Phi_3) -  \Sigma_3) \frac{\cos \theta}{\sin^2 \theta} \partial_{\phi} \nonumber \\
-\frac{1}{4} -\frac{1}{4 \sin^2 \theta} + \Sigma_3 (-i\Phi_3 ) \frac{1}{\sin^2 \theta} - (-i\Phi_3)^2 \cot^2 \theta \bigr].
\end{align}
We then act this operator on a fermion $\psi(X)$ which satisfy Eq.~(\ref{kousoku3}), and obtain the reration
\begin{equation}
(-i \hat{D})^2 \psi = -\bigl[\frac{1}{\sin \theta} \partial_{\theta} (\sin \theta \partial_{\theta}) 
+ \frac{1}{\sin^2 \theta} \partial_{\phi}^2 \bigr] \psi.
\end{equation}
The eigenvalues of the RHS operator are less than or equal to zero.
Thus the fermion components satisfying Eq.~(\ref{kousoku3}) have massless mode, while other components 
only have massive KK mode.
Note that the massless mode $\psi_0$ should be independent of $S^2$ coordinates $\theta$ and $\phi$:
\begin{equation}
\label{masslessmode}
\psi_0 = \psi(x).
\end{equation}
The existence of massless fermion may indicate the meaning of the symmetry condition; though the energy density of 
the gauge sector in the appearance of the background fields is higher than 
that of no background fields, since we have massless fermions,
it may consist a ground state as a total in the presence of fermions.
We also note that we could impose symmetry condition on fermions~\cite{Kapetanakis:1992hf,Manton:1981es}. 
In that case, we obtain the massless condition Eq.~(\ref{kousoku3}) from symmetry condition of fermion, 
and 
the solution of symmetry condition is independent 
from $S^2$ coordinate: $\psi=\psi(x)$ 
with no massive KK mode.
Therefore,
we can apply the same discussion for this case as our case  
if we 
only focus on the massless mode in our scheme.



\subsubsection{\bf A gauge symmetry and particle contents in 4D spacetime}

The symmetry conditions and the non-trivial boundary conditions substantially constrain the 
four dimensional gauge group and its representations for the particle contents.
The gauge symmetry and particle contents in 4D spacetime 
must satisfy the constraints Eq~(\ref{kousoku1}),(\ref{kousoku2}),(\ref{kousoku3})
and 
be consistent with the boundary conditions Eq~(\ref{boundary-condition4})-(\ref{boundary-condition6}).
We show the prescriptions to identify 4D gauge symmetry and particle contents below.
%


%
First, we show the prescriptions to identify 
gauge symmetry and field components which satisfy the constrants Eq~(\ref{kousoku1}),(\ref{kousoku2}),(\ref{kousoku3}). 
The gauge group $H$ 
that satisfy the constraint Eq~(\ref{kousoku1}) is identified as 
\begin{equation}
\label{H-condition}
H = C_G(U(1)_I)
\end{equation}
where $C_G(U(1)_I)$ denotes the centralizer of U(1)$_I$ in $G$~\cite{Forgacs:1979zs}. 
Note that this implies $G$ $\supset$ $H$ = $H'$ $\times$ U(1)$_I$, where $H'$ is some subgroup of $G$.
%


%
Second, the scalar field components which satisfy the constraints Eq.~(\ref{kousoku2}) are specified by the following 
prescription. 
Suppose that the adjoint representations of SU(2)$_I$ and $G$ are decomposed according to 
the embeddings SU(2)$_I$ $\supset$ U(1)$_I$ and $G$ $\supset$ $H'$ $\times$ U(1)$_I$ as 
\begin{align}
 3( \mathrm{adj} \, \mathrm{SU}(2)) & = (0(\mathrm{adj} \, \mathrm{U}(1)_R)) + (2)+(-2),
  \label{SU(2)-dec}
  \\
  \mathrm{adj} \, G
  & = (\mathrm{adj} \, H)(0)
    + 1(0(\mathrm{adj} \, \mathrm{U}(1))_R)
    + \sum_{g} h_{g}(r_{g}),
  \label{G-dec}
\end{align}
where $h_g$s denote representation of $H'$, and $r_g$s denote U(1)$_I$ charges.
The scalar components satisfying the constraints belong to 
$h_g$s whose corresponding $r_g$s in the decomposition Eq.~(\ref{G-dec}) are $\pm 2$.
%


%
Third, the fermion components which satisfy the constraints Eq.~(\ref{kousoku3}) are determined as follows~\cite{Manton:1981es}.
Let the group U(1)$_I$ be embedded into the Lorentz group SO(2) in such a way that 
the vector representation 2 of SO(2) is decomposed according to SO(2) $\supset$ U(1)$_I$ as 
\begin{equation}
\label{dec-vec}
2= (2)+(-2).
\end{equation}
This embedding specifies a decomposition of the weyl spinor representation $\sigma_6$=4 of SO(1,5) 
according to SO(1,5) $\supset$ SU(2) $\times$ SU(2) $\times$ U(1)$_I$ as
\begin{equation}
\sigma_6 = (2,1)(1) + (1,2)(-1),
\end{equation}
where SU(2) $\times$ SU(2) representations (2,1) and (1,2) correspond to 
left-handed and right-handed spinors, respectively.
We note that this decomposition corresponds to Eq.~(\ref{chiral1})(or Eq.~(\ref{chiral2})).
We then decompose $F$ according to $G$ $\supset$ $H'$ $\times$ U(1)$_I$ as
\begin{equation}
\label{dec-F}
F = \sum_f h_f(r_f).
\end{equation}
Now the fermion components satisfying the constraints are identified as  
$h_f$s whose corresponding $r_f$s in the decomposition Eq.~(\ref{dec-F}) are (1) for 
left-handed fermions and (-1) for right-handed fermions.
%


%
Finally, we show which gauge symmetry and field components remain in 4D spacetime
by surveying the consistency between the boundary conditions Eq.~(\ref{boundary-condition4})-(\ref{boundary-condition6}), 
the solutions Eq.~(\ref{kaiA}),
 and fermion massless mode Eq.~(\ref{masslessmode}).
We then apply Eq~(\ref{kaiA}) and Eq.~(\ref{masslessmode}) to Eq.~(\ref{boundary-condition4})-(\ref{boundary-condition6}), 
and obtain the parity conditions 
\begin{align}
\label{pari-con-Amu}
A_{\mu}(x) &=  P_{1(2)}A_{\mu}(x) P_{1(2)}, \\
\label{pari-con-sca1}
-\Phi_1(x) &= -P_1 (-\Phi_1(x)) P_1 , \\
\label{pari-con-sca2}
-\Phi_1(x) &= P_2 (-\Phi_1(x)) P_2, \\
\label{pari-con-sca3}
 \Phi_2(x)+ \Phi_3 \cos \theta &= -P_1 \Phi_2(x) P_1+ P_1 \Phi_3 P_1 \cos \theta, \\
 \label{pari-con-sca4}
 \Phi_2(x) - \Phi_3 \cos \theta &= P_2 \Phi_2(x) P_2 - P_2 \Phi_3 P_2 \cos \theta, \\
 \label{pari-con-psi1}
\Psi (x) &=  \gamma^5 P_1 \Psi (x), \\
\label{pari-con-psi2}
\Psi (x) &= P_2 \Psi (x).
\end{align}
We find that gauge fields, scalar fields and massless fermions in 4D spacetime should be even for 
$P_1 A_{\mu} P_1$ and $P_2 A_{\mu} P_2$; $-P_1 \Phi_{1,2} P_1 $ and $P_2 \Phi_{1,2} P_2$; 
$\gamma_5 P_1 \psi$ and $P_2 \psi$, respectively. 
$\Phi_3$ always remains since 
it 
is proportional to an U(1)$_I$ generator and commutes with $P(P')$.     
Therefore the particle contents are identified as the components which 
satisfy both the constraints Eq~(\ref{kousoku1}),(\ref{kousoku2}),(\ref{kousoku3}) and 
the parity conditions Eq Eq~(\ref{pari-con-Amu})-(\ref{pari-con-psi2}).
The gauge symmetry remained in 4D spacetime can also be identified 
by observing which components of the gauge fields remain.


\subsection{ The dimensional reduction scheme without symmetry condition}

Here we review the dimensional reduction scheme which does not apply symmetry condition of gauge field~\cite{Chiang:2010hy}.

\subsubsection{\bf Background gauge field and gauge group reduction}

For the case without symmetry condition, we consider the background gauge field 
$A^B_{\phi} \equiv {\tilde A}^B_\phi \sin\theta$ that corresponds to a Dirac monopole~\cite{background}
\begin{equation}
\label{background}
{\tilde A}^B_{\phi} = - Q \frac{\cos \theta \mp 1}{\sin \theta} ~, \quad (-: 0 \leq \theta< \frac{\pi}{2} ~, \quad +: \frac{\pi}{2} \leq \theta \leq \pi)
\end{equation}
where $Q$ is proportional to the generator of a U(1) subgroup of the original gauge group.
This background gauge field $A_{\phi}^B$ is corresponding to $\Phi_3 \cos \theta \subset A_{\phi}$  in Eq.~(\ref{kaiA}).

The background gauge field is chosen to belong to U(1)$_I$ group which is a subgroup of original gauge group $G$ such as 
\begin{equation}
G \supset G_{\rm sub} \otimes {\rm U(1)}_I,
\end{equation}
where $G_{\rm sub}$ is subgroup of $G$.
We then find that there is no massless mode for gauge field components which have non-zero U(1)$_I$ charge.
 In fact, these components acquire masses due to the background field from the term proportional to $F_{\mu \phi} F^{\mu}_{\ \phi}$
\begin{eqnarray}
&& Tr\left[
-\frac{1}{4} F_{\mu \nu}F^{\mu \nu} 
+ \frac{1}{2R^2 \sin^2 \theta} F_{\mu \phi}F^{\mu}_{\ \phi}
\right] \nonumber \\ 
&& \quad \rightarrow 
Tr\left[
-\frac{1}{4} (\partial_{\mu} A_{\nu}-\partial_{\nu} A_{\mu})(\partial^{\mu} A^{\nu}-\partial^{\nu} A^{\mu}) 
- \frac{1}{2R^2 \sin^2 \theta} [A_{\mu},A^B_{\phi}][A^{\mu},A^B_{\phi}]
\right] ~. \nonumber \\
\end{eqnarray}
For the components of $A_{\mu}$ with nonzero U(1)$_I$ charge, we have 
\begin{equation}
A_{\mu}^i Q_i+ A_{i\mu} Q^{i} \in A_{\mu} ~,
\end{equation}
where $Q_i \, (Q^i = Q_i^{\dagger})$ are generators corresponding to distinct components in Eq.~(\ref{d78}) that have nonzero U(1)$_I$ charges, and $A_{i\mu} \, (A_{\mu}^i = A_{i \mu}^{\dagger})$ are the corresponding components of $A_{\mu}$.  We then find the term
\begin{eqnarray}
\frac{1}{\sin^2 \theta}Tr[[A_{\mu},A^B_\phi][A^{\mu},A^B_\phi]]
&=& \frac{(\cos \theta \mp 1)^2}{\sin^2 \theta}   
Tr[[A_{\mu}^i Q_i+A_{i \mu} Q^i,Q][ A^{i \mu} Q_i+A_{i}^{\mu} Q^i,Q]] \nonumber \\
&=& -2 |q|^2 \frac{(\cos \theta \mp 1)^2}{\sin^2 \theta} A^{i \mu} A_{i \mu} ~,
\end{eqnarray}
where $q$ is the $Q$ charge of the relevant component.  Use of the facts that $A_{\phi}^B$ belongs to U(1)$_I$ and that $Tr[Q_i Q^i]=2$ has been made in the above equation.  A mass is thus associated with the lowest modes of those components of $A_{\mu}$ with nonzero U(1)$_I$ charges:
\begin{eqnarray}
&& \int d \Omega 
Tr\left.\left[
-\frac{1}{4}(\partial_{\mu} A_{\nu}-\partial_{\nu} A_{\mu})(\partial^{\mu} A^{\nu}-\partial^{\nu} A^{\mu})  - \frac{1}{2R^2 \sin^2 \theta} [A_{\mu},A_B][A^{\mu},A_B]
\right] \right|_{\rm lowest} \nonumber \\
&& \quad \rightarrow 
-\frac{1}{2} 
\left[ \partial_{\mu} A_{i \nu}(x) - \partial_{\nu} A_{i \mu}(x) \right]
\left[ \partial^{\mu} A^{i \nu}(x) - \partial^{\nu} A^{i \mu}(x) \right]
+ m^2_B A_{\mu}^i(x) A^{i\mu}(x) ~,
\end{eqnarray}
where the subscript `lowest' means that only the lowest KK modes are kept.  
Here the lowest KK modes of $A_{\mu}$ correspond to the term $A_{\mu}(x)/\sqrt{4 \pi}$ in the KK expansion.  In summary, any representation of $A_\mu$ carrying a nonzero U(1)$_I$ charge acquires a mass $m_B$ from the background field contribution after one integrates over the extra spatial coordinates.  More explicitly,
\begin{equation}
\label{eq:nonSMgaugeMass}
m^2_B = \frac{|q|^2}{4 \pi R^2} 
\int d \Omega \frac{(\cos \theta \mp 1)^2}{\sin^2 \theta} 
\simeq 0.39 \frac{ |q|^2}{R^2} 
\end{equation}
for the zero mode.
Therefore the gauge group $G$ is reduced to $G_{\rm sub} \otimes$U(1)$_I$ by the existence of the background gauge field.
We note that this condition is same as the case with the symmetry condition.


\subsubsection{ \bf Scalar field contents in 4D spacetime \label{sec:scalar}}

The scalar contents in 4D spacetime are obtained from the extra-dimensional components of the gauge field $\{ A_{\theta}, A_{\phi} \}$ after integrating out the extra spatial coordinates.  The kinetic term and potential term of $\{A_{\theta}, A_{\phi} \}$ are obtained from the gauge sector containing these components
\begin{eqnarray}
\label{action-scalar}
S_{\rm scalar}
&=& \int dx^4 d \Omega \Bigl(  \frac{1}{2 g^2}  Tr[ F_{\mu \theta} F^{\mu}_{\  \theta} ] + \frac{1}{2 g^2 \sin^2 \theta}  Tr[ F_{\mu \phi} F^{\mu}_{\  \phi} ] \nonumber \\
&& \qquad \qquad  -\frac{1}{2 g^2 R^2 \sin^2 \theta}  Tr[ F_{\theta \phi} F_{\theta \phi} ] \Bigr) \nonumber \\
&\rightarrow& \int dx^4 d \Omega \Bigl( \frac{1}{2 g^2} Tr[(\partial_{\mu} A_{\theta}-i[A_{\mu},A_{\theta}])^2] \nonumber \\
&& \qquad \qquad + \frac{1}{2 g^2} Tr[(\partial_{\mu} A_{\theta}-i[A_{\mu},\tilde{A}_{\phi}])^2 ] \nonumber \\
&& \qquad \qquad -\frac{1}{2 g^2 R^2} Tr \biggl[  \biggl( \frac{1}{\sin \theta} \partial_{\theta} (\sin \theta \tilde{A}_{\phi} + \sin \theta \tilde{A}^B_{\phi}) \nonumber \\
&& \qquad \qquad \ -\frac{1}{\sin \theta} \partial_{\phi} A_{\theta} 
- i[A_{\theta},\tilde{A}_{\phi}+\tilde{A}^B_{\phi}] \biggr)^2 \biggr] ~,
\end{eqnarray}
where we have taken $A_{\phi} = \tilde{A}_{\phi} \sin \theta + \tilde{A}_{\phi}^B \sin \theta$.  In the second step indicated by the arrow in Eq.~(\ref{action-scalar}), we have omitted terms which do not involve $A_{\theta}$ and $\tilde{A}_{\phi}$ from the right-hand side of the first equality.  It is known that one generally cannot obtain massless modes for physical scalar components in 4D spacetime~\cite{Maru:2006, Dohi:2010vc}.  One can see this by noting that the eigenfunction of the operator $\frac{1}{ \sin \theta } \partial_{\theta} \sin \theta$ with zero eigenvalue is not normalizable~\cite{Maru:2006}.  In other words, these fields have only KK modes.  However, an interesting feature is that it is possible to obtain a negative squared mass when taking into account the interactions between the background gauge field $\tilde{A}_{\phi}^B$ and $\{A_{\theta}, \tilde{A}_{\phi} \}$.  This happens when the component carries a nonzero U(1)$_I$ charge, as the background gauge field belongs to U(1)$_I$.  In this case, the $(\ell=1,m=1)$ modes of these real scalar components are found to have a negative squared mass in 4D spacetime.  They can be identified as the Higgs fields once they are shown to belong to the correct representation under the SM gauge group.  Here the numbers $(\ell,m)$ are the angular momentum quantum number on $S^2/Z_2$, and each KK mode is characterized by these numbers.  One can show that the $(\ell=1,m=0)$ mode has a positive squared mass and is not considered as the Higgs field.  A discussion of the KK masses with general $(\ell,m)$ will be given in Section~\ref{KKmass}~.


\subsubsection{\bf Chiral fermions in 4D spacetime \label{sec:fermions}}

We introduce fermions as the Weyl spinor fields of the 6D Lorentz group SO(1,5).  They can be written in terms of the SO(1,3) Weyl spinors as Eq.~(\ref{chiral1}) and Eq.~(\ref{chiral2}).
In general, fermions on the two-sphere do not have massless KK modes because of the positive curvature of the two-sphere.  The massless modes can be obtained by incorporating the background gauge field (\ref{background}) though, for it can cancel the contribution from the positive curvature.  In this case, the condition for obtaining a massless fermion mode is
\begin{equation}
\label{massless-condition}
Q \Psi = \pm \frac{1}{2} \Psi ~,
\end{equation}
where $Q$ comes from the background gauge field and is proportional to the U(1)$_I$ generator~\cite{background,Maru:2009wu,Dohi:2010vc}.  
We observe that the upper [lower] component on the RHS of Eq.~(\ref{chiral1}) [(\ref{chiral2})] has a massless mode for the $+$ $(-)$ sign on the RHS of Eq.~(\ref{massless-condition}).


\subsubsection{\bf The Higgs potential}

The Lagrangian for the Higgs sector is derived from the gauge sector that contains extra-dimensional components of the gauge field $\{A_{\theta}, \tilde{A}_{\phi} \}$, as given in Eq.~(\ref{action-scalar}), by considering the lowest KK modes of them.  The kinetic term and potential term are, respectively,
\begin{eqnarray}
L_{K}
&=& \frac{1}{2 g^2} \int d \Omega 
\left.
\Bigl( Tr[(\partial_{\mu} A_{\theta}-i[A_{\mu},A_{\theta}])^2] 
+ Tr[(\partial_{\mu} A_{\theta}-i[A_{\mu},\tilde{A}_{\phi}])^2 ]  \Bigr)
\right|_{\textrm{lowest}} ~, \nonumber \\ \\
V
&=& \frac{1}{2 g^2 R^2} \int d \Omega 
Tr \biggl[  \biggl( \frac{1}{\sin \theta} \partial_{\theta} (\sin \theta \tilde{A}_{\phi} 
+ \sin \theta \tilde{A}^B_{\phi}) -\frac{1}{\sin \theta} \partial_{\phi} A_{\theta} \nonumber \\
&& \qquad \qquad \qquad \qquad \left. - i[A_{\theta},\tilde{A}_{\phi}+\tilde{A}^B_{\phi}] \biggr)^2 \biggr]
\right|_{\textrm{lowest}} ~. 
\end{eqnarray}
In our model, scalar components other than the Higgs field have vanishing VEV because only the Higgs field has a negative mass-squared term, coming from the interaction with the background gauge field at tree level.  Therefore, only the Higgs field contributes to the spontaneous symmetry breaking.
Consider the $(1,1)$ mode of the $ \{ (1,2)(3,-3,3) + {\rm h.c.} \}$ representation in Eq.~(\ref{78scalar}) as argued in the previous section.  The gauge fields are given by the following KK expansions
\begin{eqnarray}
\label{expansion1}
A_{\theta} &=& - \frac{1}{\sqrt{2}} [ \Phi_1(x)  \partial_{\theta} Y_{11}^-( \theta, \phi) + \Phi_2(x) \frac{1}{\sin \theta} \partial_{\phi} Y_{11}^-( \theta, \phi) ] + \cdots ~, \\
\label{expansion2}
\tilde{A}_{\phi} &=& \frac{1}{\sqrt{2}}[ \Phi_2(x)  \partial_{\theta} Y_{11}^-( \theta,\phi)-\Phi_1(x) \frac{1}{\sin \theta}  \partial_{\phi} Y_{11}^-( \theta,\phi)] + \cdots ~,
\end{eqnarray}
where $\cdots$ represents higher KK mode terms~\cite{Maru:2009wu}.   The function $Y_{11}^- = -1/\sqrt{2} [Y_{11}+Y_{1-1}]$ is odd under $(\theta,\phi) \rightarrow (\pi/2-\theta,-\phi)$ .  We will discuss their higher KK modes and masses in the existence of the background gauge field in Section~\ref{KKmass}.  With Eqs.~(\ref{expansion1}) and (\ref{expansion2}), the kinetic term becomes
\begin{eqnarray}
L_{K}(x) = \frac{1}{2 g^2}  \Bigl( Tr[D_{\mu} \Phi_1(x) D^{\mu} \Phi_1(x)] + Tr[D_{\mu} \Phi_2(x) D^{\mu} \Phi_2(x)]  \Bigr)  ,
\end{eqnarray}
where $D_{\mu} \Phi_{1,2} = \partial_{\mu} \Phi_{1,2} -i[A_{\mu},\Phi_{1,2}]$ is the covariant derivative acting on $\Phi_{1,2}$.  The potential term, on the other hand, is
\begin{eqnarray}
V 
&=& \frac{1}{2 g^2 R^2} \int d \Omega Tr \biggl[ \biggl( -\sqrt{2} Y_{11}^- \Phi_2(x) +  Q  \nonumber \\
&& \qquad \qquad
+ \frac{i}{2}  [\Phi_1(x), \Phi_2(x)] \{ \partial_{\theta} Y_{11}^- \partial_{\theta} Y_{11}^- + \frac{1}{\sin^2 \theta} \partial_{\phi} Y_{11}^- \partial_{\phi} Y_{11}^- \} \nonumber \\
&& \qquad \qquad 
+\frac{ i}{\sqrt{2}} [\Phi_1(x), \tilde{A}^B_{\phi}] \partial_{\theta} Y_{11}^- +\frac{ i}{\sqrt{2}} [\Phi_2(x), \tilde{A}^B_{\phi}] \frac{1}{\sin \theta} \partial_{\phi} Y_{11}^-  \biggr)^2 \biggr] ~,
\nonumber \\
\end{eqnarray}
where $\partial_{\theta} (\sin \theta \tilde{A}_{\phi}^B) = Q \cos \theta$ from Eq.~(\ref{background}) is used.  Expanding the square in the trace, we get
\begin{eqnarray}
\label{potential}
V &=& \frac{1}{2 g^2 R^2} \int  d \Omega Tr
\biggl[ 2 (Y_{11}^+)^2 \Phi_2^2(x) + Q^2 \nonumber \\
&& \qquad \qquad \qquad
- \frac{1}{4} [\Phi_1(x),\Phi_2(x)]^2 \left( \partial_{\theta} Y_{11}^- \partial_{\theta} Y_{11}^- + \frac{1}{\sin^2 \theta} \partial_{\phi} Y_{11}^- \partial_{\phi} Y_{11}^- \right)^2  \nonumber \\
&& \qquad \qquad \qquad 
-\frac{1}{2} [\Phi_1(x),\tilde{A}^B_{\phi}]^2  (\partial_{\theta} Y_{11}^- )^2 -\frac{1}{2} [\Phi_2(x),\tilde{A}^B_{\phi}]^2  \left( \frac{1}{\sin \theta} \partial_{\phi} Y_{11}^- \right)^2 \nonumber \\ 
&& \qquad \qquad \qquad 
-2 i \Phi_2(x) [\Phi_1(x), \tilde{A}_{\phi}^B] 
Y_{11}^- \partial_{\theta} Y_{11}^- \nonumber \\
&& \qquad \qquad \qquad
- [\Phi_1(x),\tilde{A}_{\phi}^B] [\Phi_2(x),\tilde{A}_{\phi}^B] 
\partial_{\theta} Y_{11}^- \frac{1}{\sin \theta} \partial_{\phi} Y_{11}^- \nonumber \\
&& \qquad \qquad \qquad 
+ i Q [\Phi_1(x), \Phi_2(x)] \left( \partial_{\theta} Y_{11}^- \partial_{\theta} Y_{11}^- + \frac{1}{\sin^2 \theta} \partial_{\phi} Y_{11}^- \partial_{\phi} Y_{11}^- \right) 
~\biggr] ~, \nonumber \\
\end{eqnarray}
where terms that vanish after the $d\Omega$ integration are directly omitted.
In the end, the potential is simplified to
\begin{eqnarray}
V = \frac{1}{2 g^2 R^2} Tr \biggl[ 2 \Phi_2^2(x) + 4 \pi Q^2 - \frac{3}{10 \pi} [\Phi_1(x),\Phi_2(x)]^2 +  \frac{5i}{2} Q [\Phi_1(x), \Phi_2(x)]  \nonumber \\
+ \mu_1 [Q, \Phi_1(x)]^2 +  \mu_2 [Q, \Phi_2(x)]^2  \biggr] ~, 
\end{eqnarray}
where use of $\tilde{A}_{\phi}^B = -Q (\cos \theta \mp 1) / \sin \theta$ has been made and $\mu_1 = 1-\frac{3}{2} \ln 2$ and $\mu_2 = \frac{3}{4}(1-2\ln2)$.

We now take the following linear combination of $\Phi_1$ and $\Phi_2$ to form a complex Higgs doublet, 
\begin{eqnarray}
\label{okikae1}
\Phi(x) &=& \frac{1}{\sqrt{2}} (\Phi_1(x)+i\Phi_2(x)) ~, \\
\label{okikae2}
\Phi(x)^{\dagger} &=& \frac{1}{\sqrt{2}} (\Phi_1(x)-i\Phi_2(x)) ~.
\end{eqnarray}
It is straightforward to see that
\begin{eqnarray}
[\Phi_1(x),\Phi_2(x)] = i [\Phi(x), \Phi^{\dagger}(x)] ~.
\end{eqnarray}
The kinetic term and the Higgs potential now become 
\begin{eqnarray}
\label{kinetic-t}
L_{K} &=& \frac{1}{g^2} Tr[D_{\mu} \Phi^{\dagger}(x) D^{\mu} \Phi(x) ] ~, \\
\label{potential-t}
V &=& \frac{1}{2 g^2 R^2} Tr \biggl[ 2 \Phi_2^2(x) + 4 \pi Q^2 + \frac{3}{10 \pi} [\Phi(x),\Phi^{\dagger}(x)]^2 - \frac{5}{2}  Q [\Phi(x), \Phi^{\dagger}(x)] \nonumber \\
&& \qquad + \mu_1[Q, \Phi_1(x)]^2 + \mu_2[Q, \Phi_2(x)]^2  \biggr] ~. 
\end{eqnarray}
%


\section{The models based on our schemes \label{models}}

In this section, we show concrete models  based on the scheme introduced previous section.
We introduce the model based on SO(12) gauge symmetry for the scheme with symmetry condition and 
introduce the model based on E$_6$ gauge symmetry for the scheme without symmetry condition~\cite{Nomura:2008sx, Chiang:2010hy}.

\subsection{The SO(12) model with symmetry condtion}   
\label{SO(12)model}

Here we show a model based on a gauge group $G$=SO(12) and 
a representation $F$=32 of SO(12) for fermions,  under the scheme with symmetry condition~\cite{Nomura:2008sx}.
The choice of $G$=SO(12) and $F$=32 is motivated by the study based on CSDR which leads to  
an SO(10) $\times$ U(1) gauge theory with one generation of fermion in 4D spacetime~\cite{Chapline:1982wy} 
(for SO(12) GUT see also \cite{Rajpoot:1981it}).



\subsubsection{\bf A gauge symmetry and particle contents}

First, we show the particle contents in 4D spacetime without parities Eq.~(\ref{boundary-condition4})-(\ref{boundary-condition6}).
We assume that U(1)$_I$ is embedded into SO(12) such as 
\begin{equation}
SO(12) \supset SO(10) \times U(1)_I.
\end{equation} 
Thus we identify SO(10) $\times$ U(1)$_I$ as the gauge group which satisfy the constraints Eq~(\ref{kousoku1}), using 
Eq.~(\ref{H-condition}).
We identify the scalar components which satisfy Eq.~(\ref{kousoku2}) 
by decomposing adjoint representation of SO(12):
\begin{equation}
\label{dec66-1}
  SO(12) \supset SO(10) \times U(1)_I: 
  66 = 45(0) +1(0)+ 10(2) + 10(-2).
\end{equation}
According to the prescription below Eq.~(\ref{H-condition}) in sec.~\ref{CSDR}, 
the scalar components 10(2)+10(-2) remains in 4D spacetime.
We also identify the fermion components which satisfy Eq.~(\ref{kousoku3}) by decomposing 32 representations of SO(12) as
\begin{equation}
\label{dec32-1}
  SO(12) \supset SO(10) \times U(1)_I: 
  32 = 16(1)+ \overline{16}(-1).
\end{equation}
According to the prescription below Eq.~(\ref{G-dec}) in sec.~\ref{CSDR}, we have the fermion components as 
16(1) for a left-handed fermion and $\overline{16}$(-1) for a right-handed fermion,
 respectively, in 4D spacetime.
%


%
Next, we specify the parity assignment of $P(P')$ in order to identify the gauge symmetry and particle contents 
that actually remain in 4D spacetime.
We choose a parity assignment so as to break gauge symmetry as 
SO(12) $\supset$ SO(10) $\times$ U(1)$_I$ $\supset$ SU(5)$\times$ U(1)$_X$ $\times$ U(1)$_I$
$\supset$ SU(3) $\times$ SU(2)$_L$ $\times$ U(1)$_Y$ $\times$ U(1)$_X$ $\times$ U(1)$_I$, and to 
maintain Higgs-doublet in 4D spacetime.
The parity assignment is written in 32 dimensional spinor basis of SO(12) such as 
\begin{align}
\label{pari32}
SO(12) & \supset  SU(3) \times SU(2)_L \times U(1)_Y \times U(1)_X \times U(1)_I \nonumber \\
32
 = & (3,2)^{(+-)}(1,-1,1)+(\bar{3},2)^{(+-)}(-1,1,-1)  \nonumber \\
&+ (3,1)^{(--)}(4,1,-1)+(\bar{3},1)^{(--)}(-4,-1,1) \nonumber \\
&  + (3,1)^{(-+)}(-2,-3,-1)+(\bar{3},1)^{(-+)}(2,3,1)  \nonumber \\
&+ (1,2)^{(++)}(3,-3,-1)+(1,2)^{(++)}(-3,3,1) \nonumber \\ 
&  + (1,1)^{(--)}(6,-1,1)+(1,1)^{(--)}(-6,1,-1)   \nonumber \\
&+(1,1)^{(-+)}(0,-5,1)+(1,1)^{(-+)}(0,5,-1), 
\end{align}
where e.g. $(+,-)$ means that the parities $(P, P')$ of 
the associated components are (even, odd).
We find the gauge symmetry in 4D spacetime by surveying parity assignment for 
the gauge field.
The parity assignments of the gauge field under $A_{\mu}$ $\rightarrow$ $ PA_{\mu}P(P' A_{\mu} P')$ are: 
\begin{align}
\label{pari66-1}
66
 = & (8,1)^{(++)}(0,0,0)+(1,3)^{(++)}(0,0,0)+(1,1)^{(++)}(0,0,0) \nonumber \\
&  +(1,1)^{(++)}(0,0,0)+(1,1)^{(++)}(0,0,0) \nonumber \\
&  + \bigl[(3,2)^{(-+)}(-5,0,0)+ (\bar{3},2)^{(-+)}(5,0,0) \nonumber \\
&  +(3,2)^{(--)}(1,4,0)+ (\bar{3},2)^{(--)}(-1,-4,0) \nonumber \\ 
&  +(3,1)^{(+-)}(4,-4,0)+ (\bar{3},1)^{(+-)}(-4,4,0) \nonumber \\
&  +\underline{(3,1)^{(+-)}(-2,2,2)+ (\bar{3},1)^{(+-)}(2,-2,-2)} \nonumber \\
&  +\underline{(3,1)^{(++)}(-2,2,-2)+ (\bar{3},1)^{(++)}(2,-2,2)} \nonumber \\
&  +\underline{(1,2)^{(--)}(3,2,2)+ (1,2)^{(--)}(-3,-2,-2)} \nonumber \\
&  +\underline{(1,2)^{(-+)}(3,2,-2)+ (1,2)^{(-+)}(-3,-2,2)} \nonumber \\
&  +(1,1)^{(+-)}(6,4,0)+ (1,1)^{(+-)}(-6,-4,0) \bigr].
\end{align}
The components with 
an underline are originated from 
10(2) and 10(-2) of SO(10) $\times$ U(1)$_I$,
which do not satisfy constraints Eq.~(\ref{kousoku1}), 
and hence these components do not remain in 4D spacetime. 
Thus we have the gauge field with $(+,+)$ parity components without 
an underline in 4D spacetime,
and the gauge symmetry is SU(3) $\times$ SU(2)$_L$ $\times$ U(1)$_Y$ $\times$ U(1)$_X$ $\times$ U(1)$_I$. 
%


%
The scalar particle contents in 4D spacetime are determined by 
the parity assignment, under $\Phi_{1,2}$ $\rightarrow$ $-P \Phi_{1,2} P$ and $P' \Phi_{1,2}P'$:
\begin{align}
\label{pari66-2}
66
 = & (8,1)^{(-+)}(0,0,0)+(1,3)^{(-+)}(0,0,0)+(1,1)^{(-+)}(0,0,0) \nonumber \\
&  +(1,1)^{(-+)}(0,0,0)+(1,1)^{(-+)}(0,0,0) \nonumber \\
&  + \bigl[(3,2)^{(++)}(-5,0,0)+ (\bar{3},2)^{(++)}(5,0,0) \nonumber \\
&  +(3,2)^{(+-)}(1,4,0)+ (\bar{3},2)^{(+-)}(-1,-4,0) \nonumber \\ 
&  +(3,1)^{(--)}(4,-4,0)+ (\bar{3},1)^{(--)}(-4,4,0) \nonumber \\
&  +\underline{(3,1)^{(--)}(-2,2,2)+ (\bar{3},1)^{(--)}(2,-2,-2)} \nonumber \\
&  +\underline{(3,1)^{(-+)}(-2,2,-2)+ (\bar{3},1)^{(-+)}(2,-2,2)} \nonumber \\
&  +\underline{(1,2)^{(+-)}(3,2,2)+ (1,2)^{(+-)}(-3,-2,-2)} \nonumber \\
&  +\underline{(1,2)^{(++)}(3,2,-2)+ (1,2)^{(++)}(-3,-2,2)} \nonumber \\
&  +(1,1)^{(--)}(6,4,0)+ (1,1)^{(--)}(-6,-4,0) \bigr]. 
\end{align}
Note that the relative sign for 
the parity assignment of $P$ is different from Eq.~(\ref{pari66-1}), 
and that 
the only underlined parts satisfy the constraints Eq.~(\ref{kousoku2}).
Thus the scalar components in 4D spacetime are (1,2)(3,2,-2) and (1,2)(-3,-2,2).
%


%
We find massless fermion contents in 4D spacetime, 
by surveying the parity assignment for each components of fermion fields.
We introduce 
two types of left-handed Weyl fermions
that belong to 32 representation of SO(12), which have parity assignment 
$\psi^{(P')}$ $\rightarrow$ $ \gamma_5 P \psi^{(P')}(P' \psi^{(P')})$ 
and $\psi^{(-P')}$ $\rightarrow$ $\gamma_5 P \psi^{(-P')}(-P' \psi^{(-P')})$ respectively.
They have the parity assignment as 
\begin{align}
\label{pari32L-1}
32_L^{(P')}  
 = & \underline{(3,2)^{(--)}(1,-1,1)_L}+(\bar{3},2)^{(--)}(-1,1,-1)_L \nonumber \\
&  +\underline{(\bar{3},1)^{(+-)}(-4,-1,1)_L} + (3,1)^{(+-)}(4,1,-1)_L \nonumber \\
&  +\underline{(\bar{3},1)^{(++)}(2,3,1)_L} + (3,1)^{(++)}(-2,-3,-1)_L \nonumber \\
&  +\underline{(1,2)^{(-+)}(-3,3,1)_L} + (1,2)^{(-+)}(3,-3,-1)_L \nonumber \\ 
&  + \underline{(1,1)^{(+-)}(6,-1,1)_L}+(1,1)^{(+-)}(-6,1,-1)_L \nonumber \\
&  +\underline{(1,1)^{(++)}(0,-5,1)_L}+(1,1)^{(++)}(0,5,-1)_L,  \\
\label{pari32R-1}
32_R^{(P')} 
 = & (3,2)^{(+-)}(1,-1,1)_R+\underline{(\bar{3},2)^{(+-)}(-1,1,-1)_R} \nonumber \\ 
&  +(\bar{3},1)^{(--)}(-4,-1,1)_R + \underline{(3,1)^{(--)}(4,1,-1)_R} \nonumber \\
&  +(\bar{3},1)^{(-+)}(2,3,1)_R + \underline{(3,1)^{(-+)}(-2,-3,-1)_R} \nonumber \\
&  +(1,2)^{(++)}(-3,3,1)_R + \underline{(1,2)^{(++)}(3,-3,-1)_R} \nonumber \\ 
&  + (1,1)^{(--)}(6,-1,1)_R+ \underline{(1,1)^{(--)}(-6,1,-1)_R} \nonumber \\
&  +(1,1)^{(-+)}(0,-5,1)_R+ \underline{(1,1)^{(-+)}(0,5,-1)_R}, 
\end{align}
and 
\begin{align}
\label{pari32L-2}
32_L^{(-P')}  
 = & \underline{(3,2)^{(-+)}(1,-1,1)_L}+(\bar{3},2)^{(-+)}(-1,1,-1)_L \nonumber \\
&  +\underline{(\bar{3},1)^{(++)}(-4,-1,1)_L} + (3,1)^{(++)}(4,1,-1)_L \nonumber \\
&  +\underline{(\bar{3},1)^{(+-)}(2,3,1)_L} + (3,1)^{(+-)}(-2,-3,-1)_L \nonumber \\
&  +\underline{(1,2)^{(--)}(-3,3,1)_L} + (1,2)^{(--)}(3,-3,-1)_L \nonumber \\ 
&  + \underline{(1,1)^{(++)}(6,-1,1)_L}+(1,1)^{(++)}(-6,1,-1)_L \nonumber \\
&  +\underline{(1,1)^{(+-)}(0,-5,1)_L}+(1,1)^{(+-)}(0,5,-1)_L,  \\
\label{pari32R-2}
32_R^{(-P')} 
 = & (3,2)^{(++)}(1,-1,1)_R+\underline{(\bar{3},2)^{(++)}(-1,1,-1)_R} \nonumber \\ 
&  +(\bar{3},1)^{(-+)}(-4,-1,1)_R + \underline{(3,1)^{(-+)}(4,1,-1)_R} \nonumber \\
&  +(\bar{3},1)^{(-+)}(2,3,1)_R + \underline{(3,1)^{(-+)}(-2,-3,-1)_R} \nonumber \\
&  +(1,2)^{(+-)}(-3,3,1)_R + \underline{(1,2)^{(+-)}(3,-3,-1)_R} \nonumber \\ 
&  + (1,1)^{(-+)}(6,-1,1)_R+ \underline{(1,1)^{(-+)}(-6,1,-1)_R} \nonumber \\
&  +(1,1)^{(--)}(0,-5,1)_R+ \underline{(1,1)^{(--)}(0,5,-1)_R}, 
\end{align}
where L(R) means left-handedness(right-handedness) of fermions in 4D spacetime, and 
the underlined parts correspond to the components which satisfy constraints Eq.~(\ref{kousoku3}). 
Note the relative sign for parity assignment of $P$ between left-handed fermion and right-handed fermion, 
and that of $P'$ between 32$^{(P')}$ and 32$^{(-P')}$.   
The difference between 32$^{(P')}$ and 32$^{(-P')}$ is allowed because of the bilinear form of the fermion sector.
We thus find that 
the massless fermion components in 4D spacetime are one generation of SM-fermions with right-handed neutrino: 
$\{$(3,2)(1,-1,1)$_L$,(3,1)(4,1,-1)$_R$,(3,1)(-2,-3,-1)$_R$,(1,2)(-3,3,1)$_L$,(1,1)(-6,1,-1)$_R$,(1,1)(0,5,-1)$_R$ $\}$.



\subsubsection{\bf The Higgs sector of the model}

We analyze the Higgs-sector of our model.
The Higgs-sector $L_{\textrm{Higgs}}$ is the last two terms of Eq.~(\ref{4d-action}):
\begin{align}
L_{\textrm{Higgs}} = &- \frac{1}{2g^2} Tr[D'_{\mu}\Phi_1(x) D'^{\mu} \Phi_1(x)+D'_{\mu}\Phi_2(x) D'^{\mu} \Phi_2(x)] \nonumber \\
&- \frac{1}{2g^2} Tr[(\Phi_3+[\Phi_1(x),\Phi_2(x)])(\Phi_3+[\Phi_1(x),\Phi_2(x)])],
\end{align}
where the first term of LHS is the kinetic term of Higgs and the second term gives the Higgs potential.
We then rewrite the Higgs-sector in terms of 
genuine Higgs field in order to analyze it.
%


%
We first note that the $\Phi_i$s are written as 
\begin{equation}
\Phi_i = i \phi_i = i \phi^a_i Q_a,
\end{equation} 
where $Q_a$s are generators of gauge group SO(12), 
since $\Phi_i$s are originated from gauge fields $A_{\alpha}=iA_{\alpha}^a Q_a$;
for the gauge group generator we assume the normalization Tr($Q_aQ_b$)=-2$\delta_{ab}$. 
Note that we assumed the $-i \Phi_3$ as the generator of U(1)$_I$ embedded in SO(12),
\begin{equation}
-i \Phi_3 = Q_I.
\end{equation} 
We change the notation of the scalar fields according to Eq.~(\ref{SU(2)-dec}) such 
that, 
\begin{equation}
\phi_+ = \frac{1}{2} (\phi_1+i \phi_2), \quad
\phi_- = \frac{1}{2} (\phi_1-i \phi_2),
\end{equation}
in order to express solutions of the constraints Eq.~(\ref{kousoku2}) clearly.
The constraints Eq.~(\ref{kousoku2}) is then rewritten as 
\begin{equation}
\label{commutator}
[Q_I,\phi_+ ] = \phi_+, \qquad
[Q_I,\phi_- ] = -\phi_-.
\end{equation}
The kinetic term $L_{KE}$ and potential $V(\phi)$ term are rewritten in terms of $\phi_+$ and $\phi_-$:
\begin{align}
\label{kinetic}
L_{KE} &= -\frac{1}{g^2} Tr[D'_{\mu}\phi_+(x) D'^{\mu} \phi_-(x) ], \\
\label{potential}
V &=  -\frac{1}{2g^2} Tr[Q_I^2-4Q_I[\phi_+,\phi_- ] +4[\phi_+,\phi_- ][\phi_+,\phi_- ] ],
\end{align}
where covariant derivative $D'_{\mu}$ is $D'_{\mu}\phi_{\pm} = \partial_{\mu}\phi_{\pm} - [A_{\mu},\phi_{\pm}]$.
%


%

Next, we change the notation of SO(12) generators $Q_a$ according to decomposition Eq~(\ref{pari66-1}) such 
that 
\begin{align}
\label{generators}
Q_G =  \{ & Q_i , Q_{\alpha}, Q_Y, Q, Q_I, Q_{ax(-500)},Q^{ax(500)} \nonumber \\ 
&  Q_{ax(140)},Q^{ax(-1-40)},Q_{a(4-40)},Q^{a(-440)} \nonumber \\
&  Q_{a(-22-2)},Q^{a(2-22)},Q_{a(-222)},Q^{a(2-2-2)} \nonumber \\
&  Q_{x(322)},Q^{x(-3-2-2)},Q_{x(32-2)},Q^{x(-3-22)} \nonumber \\
&  Q(640),Q(-6-40) \}, 
\end{align} 
where the order of generators corresponds to Eq~(\ref{pari66-1}), index $i=1-8$ corresponds to SU(3) adjoint rep, 
index $\alpha=1-3$ corresponds to SU(2) adjoint rep, index $a=1-3$ corresponds to SU(3)-triplet, and 
index $x=1,2$ corresponds to SU(2)-doublet.
We write $\phi_{\pm}$ in terms of the genuine Higgs field $\phi_x$ which belongs to (1,2)(3,2,-2), such 
that
\begin{align}
\label{scalar}
\phi_+ = \phi_x Q^{x(-3-22)} \\ 
\phi_- = \phi^x Q_{x(32-2)}, 
\end{align}
where $\phi^x=(\phi_x)^{\dagger}$.  
We also write gauge field $A_{\mu}(x)$ in terms of $Q$s in Eq.~(\ref{generators}) as 
\begin{equation}
\label{gauge}
A_{\mu}(x) = i(A_{\mu}^i Q_i+A_{\mu}^{\alpha} Q_{\alpha}+B_{\mu} Q_Y+C_{\mu} Q+E_{\mu} Q_I).
\end{equation}
We then need commutation relations of $Q^{x(-3-22)}$, $Q_{x(32-2)}$, $Q_{\alpha}$, $Q_Y$, $Q$ and $Q_I$
in order to analyze 
the Higgs sector; we summarized them in Table~\ref{commutators}.
\begin{center}
\begin{table}
\begin{tabular}{lll} \hline
& \multicolumn{2}{l} [$Q^{x(-3-22)}$,$Q_{y(32-2)}$] = 
$-\sqrt{\frac{3}{10}}$ $\delta^x_y$ $Q_Y$ + $-\sqrt{\frac{1}{5}}$ $\delta^x_y$ $Q$
+$\delta^x_y$ $Q_I$ +$\frac{1}{\sqrt{2}}$ $(\sigma^*_{\alpha})^x_y$ $Q_{\alpha}$  \\
&
[$Q_{\alpha}$,$Q_x$] = $-\frac{1}{\sqrt{2}}$ $(\sigma_{\alpha})_x^y$ $Q_y$ \qquad \qquad & 
[$Q_{\alpha}$,$Q^x$] = $\frac{1}{\sqrt{2}}$ $(\sigma^*_{\alpha})^x_y$ $Q^y$ \\ 
&
[$Q_x$,$Q_y$]=0 &
[$Q_Y$,$Q^x$]= $-\sqrt{\frac{3}{10}}$ $Q^x$ \\
&
[$Q$,$Q^x$]= $-\sqrt{\frac{1}{5}}$ $Q^x$ &
[$Q_I$,$Q^x$] = $Q^x$ \\ \hline 
\end{tabular}
\caption{commutation relations of $Q^{x(-3-22)}$, $Q_{x(32-2)}$, $Q_{\alpha}$, $Q_Y$, $Q$ and $Q_I$}
\label{commutators}
\end{table}
\end{center}
Finally, we obtain the Higgs sector with genuine Higgs field
by substituting Eq.~(\ref{scalar})-(\ref{gauge}) into Eq.~(\ref{kinetic}, \ref{potential})
and rescaling the fields $\phi \rightarrow g/\sqrt{2} \phi$ and $A_{\mu} \rightarrow g A_{\mu}$, 
and the couplings $\sqrt{2}g=g_2$ and $\sqrt{6/5} g = g_Y$, 
\begin{equation}
L_{Higgs} = |D_{\mu} \phi_x|^2 - V(\phi),
\end{equation}
where the covariant derivative $D_{\mu} \phi_x$ and potential $V(\phi)$ are 
\begin{align}
D_{\mu} \phi_x &= \partial_{\mu} \phi_x + i g_2 \frac{1}{2} (\sigma_{\alpha})_x^y A_{\alpha \mu} \phi_y 
+ i g_Y \frac{1}{2} B_{\mu} \phi_x + i \sqrt{\frac{1}{5}} g C_{\mu} \phi_x - ig E_{\mu} \phi_x , \\
V &= -\frac{2}{R^2} \phi^x \phi_x + \frac{3g^2}{2} (\phi^x \phi_x)^2,
\end{align}
respectively.
Notice that we explicitly write radius $R$ of $S^2$ in the Higgs potential, 
and that we omitted the constant term in the Higgs potential.
We note that the SU(2)$_L$ $\times$ U(1)$_Y$ parts of the Higgs sector has the same form as the SM Higgs sector. 
Therefore we obtain the electroweak symmetry breaking SU(2)$_L$ $\times$ U(1)$Y$ $\rightarrow$ U(1)$_{EM}$. 
The Higgs field $\phi^x$ acquires vaccume expectation value(VEV) as 
\begin{align}
<\phi> &= \frac{1}{\sqrt{2}} \begin{pmatrix} 0 \\ v \end{pmatrix}, \\
v &= \sqrt{\frac{4}{3}} \frac{1}{g R},
\end{align} 
and W boson mass $m_W$ and Higgs mass $m_H$ are given in terms of radius $R$  
\begin{align}
&  m_{W} = g_2 \frac{v}{2} = \sqrt{\frac{2}{3}} \frac{1}{R}, \\
&  m_H = \sqrt{3}g v = \sqrt{4} \frac{1}{R}.
\end{align} 
The ratio between $m_W$ and $m_H$ is predicted 
\begin{equation}
\frac{m_H}{m_W} = \sqrt{6}.
\end{equation}
We thus find $m_H \sim 196 $GeV in this model. 
We also find the Weinberg angle is given by 
\begin{eqnarray}
\sin^2 \theta_W &=& \frac{g_Y^2}{g_2^2+g_Y^2} \nonumber \\
&=& \frac{6/5}{2+6/5} \nonumber \\
&=& \frac{3}{8}
\end{eqnarray}
which is same as SU(5) GUT case.


\subsection{The E$_6$ model without symmetry condition}

Here we show a model based on a gauge group $G$=E$_6$ and 
a representation 27 of E$_6$ for fermions,  under the scheme without symmetry condition~\cite{Chiang:2010hy}.

\subsubsection{\bf Gauge group reduction}

We consider the following gauge group reduction
\begin{eqnarray}
\label{group-red}
E_6 &\supset& SO(10) \times U(1)_I \nonumber \\
&\supset& SU(5) \times U(1)_X \times U(1)_I \nonumber \\
&\supset& SU(3) \times SU(2) \times U(1)_Y \times U(1)_X \times U(1)_I ~.
\end{eqnarray}
The background gauge field in Eq.~(\ref{background}) is chosen to belong to the U(1)$_I$ group.  This choice is needed in order to obtain chiral SM fermions in 4D spacetime to be discussed later.  There are two other symmetry reduction schemes.  One can prove that the results in those two schemes are effectively the same as the one considered here once we require the correct U(1) combinations for the hypercharge and the background field.

We then impose the parity assignments with respect to the fixed points, Eqs.~(\ref{boundary-condition1})-(\ref{boundary-condition6}).  The parity assignments for the fundamental representation of E$_6$ is chosen to be
\begin{eqnarray}
\label{d27}
{\bf 27} &=& (1,2)(-3,-2,-2)^{(+,+)}+(1,2)(3,2,-2)^{(-,-)}+(1,2)(-3,3,1)^{(+,-)}  \nonumber \\ 
&& + (1,1)(6,-1,1)^{(+,+)}+(1,1)(0,-5,1)^{(-,-)}+(1,1)(0,0,4)^{(-,+)}  \nonumber \\
&& + (3,2)(1,-1,1)^{(-,+)}+(3,1)(-2,2,-2)^{(+,-)} + (\bar{3},1)(-4,-1,1)^{(+,+)} \nonumber \\
&& +(\bar{3},1)(2,3,1)^{(+,+)}+(\bar{3},1)(2,-2,2)^{(-,+)}, 
\end{eqnarray}
where, for example, $(+,-)$ means that the parities under $P_1$ and $P_2$ are (even,odd).  By the requirement of consistency, we find that the components of $A_{\mu}$ in the adjoint representation have the parities under $A_{\mu} \rightarrow P_1 A_{\mu} P_1$ $(P_2 A_{\mu} P_2)$ as follows:
\begin{eqnarray}
\label{d78}
{\bf 78}|_{A_{\mu}} &=& \underline{(8,1)(0,0,0)^{(+,+)}+(1,3)(0,0,0)^{(+,+)}} \nonumber \\
&& + \underline{(1,1)(0,0,0)^{(+,+)} +(1,1)(0,0,0)^{(+,+)}+(1,1)(0,0,0)^{(+,+)}}  \nonumber \\
&& + (3,2)(-5,0,0)^{(-,+)}+(\bar{3},2)(5,0,0)^{(-,+)}  \nonumber \\
&& ++   (3,2)(1,4,0)^{(+,-)}+ (\bar{3},2)(-1,-4,0)^{(+,-)} \nonumber \\
&&  +(3,1)(4,-4,0)^{(-,-)}+(\bar{3},1)(-4,4,0)^{(-,-)} \nonumber \\
&& + +(1,1)(-6,-4,0)^{(-,-)}+(1,1)(6,4,0)^{(-,-)} \nonumber \\
&& +  (3,2)(1,-1,-3)^{(+,+)}+  (\bar{3},2)(-1,1,3)^{(+,+)}   \nonumber \\
&& +(3,1)(4,1,3)^{(-,+)}+(\bar{3},1)(-4,-1,-3)^{(-,+)} \nonumber \\
&& + (3,1)(-2,-3,3)^{(+,-)}+ (\bar{3},1)(2,3,-3)^{(+,-)}    \nonumber \\
&& +(1,2)(-3,3,-3)^{(-,-)}+(1,2)(3,-3,3)^{(-,-)} \nonumber \\
&& +(1,1)(-6,1,3)^{(-,+)}+(1,1)(6,-1,-3)^{(-,+)}    \nonumber \\
&& +(1,1)(0,-5,-3)^{(+,-)}+(1,1)(0,5,3)^{(+,-)},
\end{eqnarray}
where the underlined components correspond to the adjoint representations of SU(3) $\times$ SU(2) $\times$ U(1)$_Y$ $\times$ U(1)$_X$ $\times$ U(1)$_I$, respectively.  We note that the components with parity $(+,+)$ can have massless zero modes in 4D spacetime.  Such components include the adjoint representations of SU(3) $\times$ SU(2) $\times$ U(1)$^3$, $(3,2)(1,-1,-3)$ and its conjugate.  The latter components seem problematic since they do not appear in the low-energy spectrum due to non-zero U(1)$_I$ charge.


\subsubsection{\bf Scalar field contents in 4D spacetime}

With the parity assignments with respect to the fixed points, Eqs.~(\ref{boundary-condition2}) and (\ref{boundary-condition5}), we have for the $A_{\theta}$ and $A_{\phi}$ fields
\begin{eqnarray}
\label{78scalar}
{\bf 78}|_{A_{\theta,\phi}} & = &  (8,1)(0,0,0)^{(-,-)}+(1,3)(0,0,0)^{(-,-)} \nonumber \\
&& +(1,1)(0,0,0)^{(-,-)} +(1,1)(0,0,0)^{(-,-)}+(1,1)(0,0,0)^{(-,-)}  \nonumber \\
&& + (3,2)(-5,0,0)^{(+,-)}+(\bar{3},2)(5,0,0)^{(+,-)}  \nonumber \\
&& +   (3,2)(1,4,0)^{(-,+)}+ (\bar{3},2)(-1,-4,0)^{(-,+)} \nonumber \\
&&  +(3,1)(4,-4,0)^{(+,+)}+(\bar{3},1)(-4,4,0)^{(+,+)} \nonumber \\
&& +(1,1)(-6,-4,0)^{(+,+)}+(1,1)(6,4,0)^{(+,+)} \nonumber \\
&& +  (3,2)(1,-1,-3)^{(-,-)}+  (\bar{3},2)(-1,1,3)^{(-,-)}   \nonumber \\
&& +(3,1)(4,1,3)^{(+,-)}+(\bar{3},1)(-4,-1,-3)^{(+,-)} \nonumber \\
&& + (3,1)(-2,-3,3)^{(-,+)}+ (\bar{3},1)(2,3,-3)^{(-,+)}    \nonumber \\
&& + +(1,2)(-3,3,-3)^{(+,+)}+(1,2)(3,-3,3)^{(+,+)} \nonumber \\
&& +(1,1)(-6,1,3)^{(+,-)}+(1,1)(6,-1,-3)^{(+,-)} \nonumber \\
&& +(1,1)(0,-5,-3)^{(-,+)}+(1,1)(0,5,3)^{(-,+)} ~.
\end{eqnarray}
Components with $(+,-)$ or $(-,+)$ parity do not have KK modes since they are odd under $\phi \rightarrow \phi+2\pi$ and the KK modes of gauge field are specified by integer angular momentum quantum numbers $\ell$ and $m$ on the two-sphere.  We then concentrate on the components which have either $(+,+)$ or $(-,-)$ parity and nonzero U(1)$_I$ charges as the candidate for the Higgs field.  These include $\{ (1,2)(3,-3,3) + {\rm h.c.} \}$ and $\{(3,2)(1,-1,-3) + {\rm h.c.} \}$ with parities $(+,+)$ and $(-,-)$, respectively.  The representations $(1,2)(-3,3,-3)$ and $(1,2)(3,-3,3)$ have the correct quantum numbers for the SM Higgs doublet.  Therefore, we identify the $(1,1)$ mode of these components as the SM Higgs fields in 4D spacetime.


\subsubsection{Chiral fermion contents in 4D spacetime}

In our model, we choose the fermions as the Weyl fermions $\Psi_-$ belonging to the {\bf 27} representation of E$_6$.  The {\bf 27} representation is decomposed as in Eq.~(\ref{d27}) under the group reduction, Eq.~(\ref{group-red}).  In this decomposition, we find that our choice of the background gauge field of U(1)$_I$ is suitable for obtaining massless fermions since all such components have U(1)$_I$ charge 1.  In the fundemantal representation, the U(1)$_I$ generator is
\begin{equation}
 Q_I   = \frac{1}{6} \mbox{diag}
(-2,-2,-2,-2,1,1,1,1,4,1,1,1,1,1,1,-2,-2,-2,1,1,1,1,1,1,-2,-2,-2) ~,
\nonumber \\
\end{equation}
according to the decomposition Eq.~(\ref{d27}).  By identifying $Q=3Q_I$, we readily obtain the condition
\begin{equation}
Q \Psi_- = \frac{1}{2} \Psi_-.
\end{equation}
Therefore, the chiral fermions $\psi_L$ in 4D spacetime have zero modes.

Next, we consider the parity assignments for the fermions with respect to the fixed points of $S^2/Z_2$.  The boundary conditions are given by Eqs.~(\ref{boundary-condition3}) and (\ref{boundary-condition6}).  It turns out that four ${\bf 27}$ fermion copies with different boundary conditions are needed in order to obtain an entire generation of massless SM fermions.  They are denoted by $\Psi^{(1,2,3,4)}$ with the following parity assignments
\begin{eqnarray}
\Psi_{\pm}^{(i)} (x,\pi-\theta,-\phi) 
&=& \xi \gamma_5 P_1 \Psi_{\pm}^{(i)}(x,\theta,\phi) ~, \\
\Psi_{\pm}^{(i)} (x,\pi-\theta,2\pi-\phi) 
&=& \eta \gamma_5 P_2 \Psi_{\pm}^{(i)}(x,\theta,\phi) ~,
\end{eqnarray}
where $\gamma_5$ is the chirality operator, and $(\xi,\eta) = (+,+)$, $(-,-)$, $(+,-)$ and $(-,+)$ for $i = 1,2,3,4$, respectively.  From these fermions we find that $\psi_{1,2,3,4}$ have the parity assignments
\begin{eqnarray}
{\bf 27}_{\psi_L^{(1)}} &=& (1,2)(-3,-2,-2)^{(-,-)}+(1,2)(3,2,-2)^{(+,+)}+(1,2)(-3,3,1)^{(-,+)}  \nonumber \\ 
&& + (1,1)(6,-1,1)^{(-,-)}+ \underline{(1,1)(0,-5,1)^{(+,+)}}+(1,1)(0,0,4)^{(+,-)}  \nonumber \\
&& + (3,2)(1,-1,1)^{(+,-)}+(3,1)(-2,2,-2)^{(-,+)} + (\bar{3},1)(-4,-1,1)^{(-,-)} \nonumber \\
&& +(\bar{3},1)(2,3,1)^{(-,-)}+(\bar{3},1)(2,-2,2)^{(+,-)} \\
{\bf 27}_{\psi_L^{(2)}} &=& (1,2)(-3,-2,-2)^{(+,+)}+(1,2)(3,2,-2)^{(-,-)}+(1,2)(-3,3,1)^{(+,-)}  \nonumber \\ 
&& + \underline{(1,1)(6,-1,1)^{(+,+)} } + (1,1)(0,-5,1)^{(-,-)}+(1,1)(0,0,4)^{(-,+)}  \nonumber \\
&& + (3,2)(1,-1,1)^{(-,+)}+(3,1)(-2,2,-2)^{(+,-)} + \underline{ (\bar{3},1)(-4,-1,1)^{(+,+)} } \nonumber \\
&& + \underline{(\bar{3},1)(2,3,1)^{(+,+)} }+(\bar{3},1)(2,-2,2)^{(-,+)} \\
{\bf 27}_{\psi_L^{(3)}} &=& (1,2)(-3,-2,-2)^{(-,+)}+(1,2)(3,2,-2)^{(+,-)}+(1,2)(-3,3,1)^{(-,-)}  \nonumber \\ 
&& + (1,1)(6,-1,1)^{(-,+)}+(1,1)(0,-5,1)^{(+,-)}+(1,1)(0,0,4)^{(+,+)}  \nonumber \\
&& + \underline{ (3,2)(1,-1,1)^{(+,+)} } +(3,1)(-2,2,-2)^{(-,-)} + (\bar{3},1)(-4,-1,1)^{(-,+)} \nonumber \\
&& +(\bar{3},1)(2,3,1)^{(-,+)}+(\bar{3},1)(2,-2,2)^{(+,+)} \\
{\bf 27}_{\psi_L^{(4)}} &=& (1,2)(-3,-2,-2)^{(+,-)}+(1,2)(3,2,-2)^{(-,+)}+ \underline{ (1,2)(-3,3,1)^{(+,+)} }  \nonumber \\ 
&& + (1,1)(6,-1,1)^{(+,-)}+(1,1)(0,-5,1)^{(-,+)}+(1,1)(0,0,4)^{(-,-)}  \nonumber \\
&& + (3,2)(1,-1,1)^{(-,-)}+(3,1)(-2,2,-2)^{(+,+)} + (\bar{3},1)(-4,-1,1)^{(+,-)} \nonumber \\
&& +(\bar{3},1)(2,3,1)^{(+,-)}+(\bar{3},1)(2,-2,2)^{(-,-)} ~, 
\end{eqnarray}
where the underlined components have even parities and U(1)$_I$ charge 1.  
One can readily identify one generation of SM fermions, including a right-handed neutrino, as the zero modes of these components.

A long-standing problem in the gauge-Higgs unification framework is the Yukawa couplings of the Higgs boson to the matter fields.  Here we discuss about the Yukawa couplings in our model.  As mentioned before, the SM Higgs is the $(\ell=1,|m|=1)$ KK mode of the extra-spatial component of the gauge field, the Yukawa term at tree level has the following form
\begin{equation}
\label{yukawa}
L_{\textrm{Yukawa}} \supset \bar{\psi}_L^{00} \Phi^{11} \psi_R^{\ell 1} + \bar{\psi}_L^{\ell1} \Phi^{11} \psi_R^{00}+ {\rm h.c.} ~,
\end{equation}
where $\psi^{\ell m}$s are the fermionic KK modes with the $(l=0,m=0)$ modes appearing as the chiral fermions and $\Phi^{11}$ denotes the SM Higgs field.  We here identify the left-handed fermionic zero modes as SU(2) doublets and the right-handed fermionic zero modes as SU(2) singlets, as in the SM.  Therefore, the $(\ell,|m|=1)$ modes and the $(\ell=0,|m|=0)$ modes mix after spontaneous symmetry breaking.  One needs to diagonalize the mass terms to obtain physical eigenstates.  The Yukawa couplings in our model are thus more complicated than other gauge-Higgs unification models.  However, the difficulty of obtaining the realistic fermion mass spectrum remains since the Yukawa couplings all arise from gauge interactions.  One way to get realistic Yukawa couplings is to introduce SM fermions localized at an orbifold fixed point and make use of nonlocal interactions with Wilson lines \cite{Csaki:2002ur}.  Another possible solution is to consider fermions in 6D spacetime belonging to a higher dimensional representation of the original E$_6$ gauge group, rendering more than one generation of SM fermions.  In that case, mixing among generations will be obtained from gauge interactions and is given by Clebsch-Gordan coefficients.  We expect that realistic Yukawa couplings could be obtained using a combination of these methods.  A detailed analysis of this issue is beyond the scope of the paper and left for a future work.


\subsubsection{\bf Higgs potential of the model}

Here we analyze the Higgs potential for the E$_6$ model.
To further simplify the Higgs potential, we need to find out the algebra of the gauge group generators.  Note that the E$_6$ generators are chosen according to the decomposition of the adjoint representation given in Eq.~(\ref{d78})
\begin{eqnarray}
&&\{ Q_i, Q_{\alpha}, Q_Y, Q_X, Q_I, \nonumber \\
&& \quad Q_{ax (-5,0,0)}, Q^{ax(5,0,0)}, Q_{ax(1,4,0)}, Q^{ax(-1,-4,0)}, \nonumber \\
&& \quad Q_{a(4,-4,0)}, Q^{a(-4,4,0)}, Q_{(-6,-4,0)}, Q_{(6,4,0)}, \nonumber \\
&& \quad Q_{ax(1,-1,-3)}, Q^{ax(-1,1,3)}, Q_{a(4,1,3)}, Q^{a(-4,-1,-3)}, \nonumber \\
&& \quad Q_{a(-2,-3,3)}, Q^{a(2,3,-3)}, Q_{x(3,-3,3)}, Q^{x(-3,3,-3)}, \nonumber \\
&& \quad Q_{(-6,1,3)}, Q_{(6,-1,-3)}, Q_{(0,-5,-3)},Q_{(0,5,3)}  \} ~,
\end{eqnarray}
where the generators are listed in the corresponding order of the terms in Eq.~(\ref{d78}) and the indices
\begin{eqnarray}
\label{generators}
&& i=1,...,8: \textrm{SU(3) adj rep index} \Rightarrow Q_i:  \textrm{SU(3) generators} ~, \\
&& \alpha=1,2,3: \textrm{SU(2) adj rep index} \Rightarrow Q_{\alpha}: \textrm{SU(2) generators} ~,  \\
&& Q_{X,Y,Z}: \textrm{$U(1)_{X,Y,Z}$ generators} ~, \\
&& x=1,2: \textrm{SU(2) doublet index} ~, \\
&& a=1,2,3: \textrm{SU(3) color index} ~.
\end{eqnarray}
Here we take the standard normalization for generators, $Tr[Q Q^{\dagger}] = 2$. 
The Higgs fields are in the representations of $(1,2)(3,-3,3)$ and $(1,2)(-3,3,-3)$.  We write
\begin{equation}
\label{Higgs}
\Phi(x) = \phi^{x} Q_{x(3,-3,3)} \quad (\Phi^{\dagger}(x) = \phi_x Q^{x(-3,3,-3)}) ~.
\end{equation}
Likewise, the gauge field $A_{\mu}(x)$ in terms of the $Q$'s in Eq.~(\ref{generators}) is
\begin{equation}
\label{gauge}
 A_{\mu}(x) = A_{\mu}^i Q_i+A_{\mu}^{\alpha} Q_{\alpha}+B_{\mu} Q_Y+C_{\mu} Q_X+E_{\mu} Q_I.
\end{equation}
The commutation relations between the generators $Q_{\alpha}$, $Q_{X,Y,Z}$, $Q_{x(3,-3,3)}$ and $Q^{x(-3,3,-3)}$ are summarized in Table.~\ref{commutators}.

\begin{center}
\begin{table}
\begin{tabular}{lll}
\hline\hline
& \multicolumn{2}{c}{
$\left[ Q_{x(3,-3,3)},Q^{y(-3,3,-3)} \right] 
= \frac{1}{2} \delta_x^y Q_I-\frac{1}{2} \sqrt{\frac{3}{5}} \delta_x^y Q_X
+ \frac{1}{\sqrt{10}} \delta_x^y Q_Y + \frac{1}{\sqrt{6}} (\sigma_{\alpha})^y_x Q_{\alpha}$}  \\
&
$\left[ Q_{\alpha},Q_{x(3,-3,3)} \right] = \frac{1}{\sqrt{6}} (\sigma_{\alpha} )^y_x Q_{y(3,-3,3)}$ \qquad \qquad & 
$\left[ Q_{\alpha},Q^{ x(-3,3,-3) } \right] = - \frac{1}{\sqrt{6}} (\sigma_{\alpha}^* )^y_x Q^{y(-3,3,-3) }$  \\ 
&
$\left[ Q_{x(3,-3,3)},Q_{y(3,-3,3)} \right] = 0$ &
$\left[ Q_I, Q_{x(3,-3,3)} \right] = \frac{1}{2} Q_{x(3,-3,3)}$  \\
&
$\left[ Q_X, Q_{x(3,-3,3)} \right] = -\frac{1}{2} \sqrt{\frac{3}{5}} Q_{x(3,-3,3)}$ &
$\left[ Q_Y, Q_{x(3,-3,3)} \right] = \frac{1}{\sqrt{10}} Q_{x(3,-3,3)}$ \\
\hline\hline 
\end{tabular}
\caption{Commutation relations of $Q_{\alpha}$, $Q_{X,Y,Z}$, $Q_{x(3,-3,3)}$ and $Q^{x(-3,3,-3)}$, where $\sigma_i$ are the Pauli matrices.}
\label{commutators}
\end{table}
\end{center}

Finally, we obtain the Lagrangian associated with the Higgs field by applying Eqs.~(\ref{Higgs}, \ref{gauge}) to Eqs.~(\ref{kinetic-t}, \ref{potential-t}) and carrying out the trace.  Furthermore, to obtain the canonical form of kinetic terms, the Higgs field, the gauge field, and the gauge coupling need to be rescaled in the following way:
\begin{eqnarray}
\label{notation}
&& \phi \rightarrow \frac{g}{\sqrt{2}} \phi \\
&& A_{\mu} \rightarrow \frac{g}{R}A_{\mu} \\
&& \frac{g}{\sqrt{6 \pi R^2}} = g_2 ~,
\end{eqnarray}
where $g_2$ denotes the SU(2) gauge coupling.  The Higgs sector is then given by
\begin{equation}
{\cal L}_{\rm Higgs} = |D_{\mu} \phi|^2 - V(\phi)
\end{equation}
where
\begin{eqnarray}
\label{cova}
D_{\mu} \phi &=& 
\left[ \partial_{\mu} + i g_2 \frac{\sigma_{\alpha}}{2}  A_{\alpha \mu}
+ ig \frac{1}{\sqrt{40 \pi R^2}} B_{\mu} 
- ig \frac{1}{2 } \sqrt{\frac{3}{20 \pi R^2}} C_{\mu}
+ i g \frac{1}{2 \sqrt{4 \pi R^2}} E_{\mu} \right] \phi ~, \nonumber \\ \\
\label{H-potential}
V &=&  -\frac{\chi}{8 R^2} \phi^{\dagger} \phi + \frac{3 g^2}{40 \pi R^2} \left(\phi^{\dagger} \phi \right)^2 ~,
\end{eqnarray}
where $\chi=7+9\mu_1+9\mu_2$.
We have omitted the constant term in the Higgs potential.  Comparing the potential derived above with the standard form $\mu^2\phi^\dagger\phi + \lambda (\phi^\dagger\phi)^2$ in the SM, we see that the model has a tree-level $\mu^2$ term that is negative and proportional to $R^{-2}$.  Moreover, the quartic coupling $\lambda = 3 g^2 / (40 \pi R^2)$ is related to the 6D gauge coupling $g$ and grants perturbative calculations because it is about $0.16$, using the value of $R$ to be extracted in the next section.  Therefore, the order parameter in this model is controlled by a single parameter $R$, the compactification scale.

In fact, the $(1,1)$ mode of the $\{(3,2)(1,-1,-3) + {\rm h.c.}\}$ representation also has a negative squared mass term because it has the same $Q_I$ charge as the $\{(1,2)(3,-3,3) + {\rm h.c.}\}$ representation.  Therefore, it would induce not only electroweak symmetry breaking but also color symmetry breaking.  This undesirable feature can be cured by adding brane terms
\begin{eqnarray}
\frac{\alpha}{R^2\sin^2\theta}
\left( F_{\theta\phi}^a F^{a\theta\phi} \right)^2
\delta\left( \theta-\frac{\pi}2 \right)
\left[
\delta(\phi)
+ \delta(\phi-\pi)
\right] ~,
\end{eqnarray}
where $a$ denotes the group index of the $\{(3,2)(1,-1,-3) + {\rm h.c.}\}$ representation.  These brane terms preserve the $Z'_2$ symmetry which corresponds to the symmetry under the transformation $(\phi \rightarrow \phi+\pi)$.  With an appropriate choice of the dimensionless constant $\alpha$, the squared mass of the $(1,1)$ can be lifted to become positive and sufficiently large.

Due to a negative mass term, the Higgs potential in Eq.~(\ref{H-potential}) can induce the spontaneous symmetry breakdown: SU(2) $\times$ U(1)$_Y$ $\rightarrow$ U(1)$_{\rm EM}$ in the SM.  The Higgs field acquires a vaccum expectation value (VEV)
\begin{eqnarray}
\langle \phi \rangle =
\frac{1}{\sqrt{2}} \begin{pmatrix} 0  \\ v \end{pmatrix} \mbox{ with }
v = \sqrt{\frac{5 \pi \chi}{3}} \frac{1}{g} 
\simeq \frac{4.6}{g} ~.
\end{eqnarray}
One immedialtey finds that the $W$ boson mass
\begin{equation}
m_W = \frac{g_2}{2} v = \frac{1}{6}\sqrt{ \frac{5 \chi}{2} } \frac{1}{R} 
\simeq \frac{0.53}{R},
\end{equation}
from which the compactification scale $R^{-1} \simeq 152$ GeV is inferred.  Moreover, the Higgs boson mass at the tree level is
\begin{equation}
m_H = \sqrt{\frac{3}{20 \pi}} \frac{g v}{R} = 3 \sqrt{\frac{2}{5}} m_W 
= \frac{\sqrt{\chi}}{2}  \frac{1}{R} ~,
\end{equation}
which is about $152$ GeV, numerically very close to the compactification scale.  Since the hypercharge of the Higgs field is $1/2$, the U(1)$_Y$ gauge coupling is derived from Eq.~(\ref{cova}) as
\begin{equation}
g_Y = \frac{g}{\sqrt{10 \pi R^2}} ~.
\end{equation}
The Weinberg angle is thus given by 
\begin{eqnarray}
\sin^2 \theta_W = \frac{g_Y^2}{g_2^2+g_Y^2}
= \frac{3}{8} ~,
\end{eqnarray}
and the $Z$ boson mass
\begin{eqnarray}
m_Z = \frac{m_W}{\cos \theta_W} = m_W \sqrt{\frac{8}{5}} ~,
\end{eqnarray}
both at the tree level.  These relations are the same as the SU(5) GUT at the unification scale.  This is not surprising because this part only depends on the group structure.


\subsubsection{KK mode spectrum of each field \label{KKmass}}

Since we did not impose symmetry condition, we have KK mode for each field in this model.  
Here we show KK mass spectrum under the existence of background field for our E$_6$ model.
The masses are basically conrtrolled by the compactification radius $R$ of the two-sphere.  They receive two kinds of contributions: one arising from the angular momentum in the $S^2$ space, and the other coming from the interactions with the background field.

The KK masses for fermions have been given in Refs.~\cite{background, Maru:2009wu, Dohi:2010vc}.  We give them in terms of our notation here:
\begin{eqnarray}
\label{KKmass-fermion}
M_{\ell m}^{KK}(\psi_L) = \frac{1}{R} \sqrt{\ell(\ell+1)-\frac{4q^2-1}{4} } ~,
\end{eqnarray}
where $q$ is proportional to the U(1)$_I$ charge of fermion and determined by the action of $Q=3Q_I$ on fermions as $Q \Psi = q \Psi = 3q_I \Psi$.   Note that the mass does not depend on the quantum number $m$.  The lightest KK mass, corresponding to $\ell = 1$ and $q_I = 1/6$, is about 214 GeV at the tree level.
The range of $\ell$ is 
\begin{equation}
\frac{2q \pm 1}{2} \leq \ell \qquad  (+: \ \textrm{for} \ \psi_{R(L)} \ \textrm{in} \ \Psi_{+(-)},  \quad - : \ \textrm{for} \ \psi_{L(R)} \ \textrm{in} \ \Psi_{-(+)} ) ~. 
\end{equation}
We thus can have zero mode for $Q \Psi = \pm \frac{1}{2} \Psi$, where this condition is given in Eq.~(\ref{massless-condition}).

For the 4D gauge field $A_{\mu}$, its kinetic term and KK mass term are obtained from the terms
\begin{equation}
\label{FF}
L=\int d \Omega Tr \biggl[ -\frac{1}{4}F_{\mu \nu} + \frac{1}{2 R^2} F_{\mu \theta} F^{\mu}_{\ \theta}+\frac{1}{2 R^2 \sin^2 \theta} F_{\mu \phi} F^{\mu}_{\ \phi} \biggr] ~.
\end{equation}
Taking terms quadratic in $A_{\mu}$, we get 
\begin{eqnarray}
L_{\rm quad} 
&=& \int d \Omega Tr \biggl[ -\frac{1}{4}(\partial_{\mu} A_{\nu}-\partial_{\nu}A_{\mu} )(\partial^{\mu}A^{\nu}-\partial^{\nu}A^{\mu} ) 
+\frac{1}{2 R^2} \partial_{\theta} A_{\mu} \partial_{\theta} A^{\mu} 
\nonumber \\
&& \qquad \qquad 
+ \frac{1}{2 R^2 \sin^2 \theta} \partial_{\phi} A_{\mu} \partial_{\phi} A^{\mu} -\frac{1}{2 R^2} [A_{\mu}, \tilde{A}_{\phi}^B][A^{\mu},\tilde{A}_{\phi}^B] \biggr] ~,
\end{eqnarray}
where $\tilde{A}^B_{\phi}$ is the background gauge field given in Eq.~(\ref{background}).  The KK expansion of $A_{\mu}$ is
\begin{equation}
A_{\mu} = \sum_{\ell m} A_{\mu}^{\ell m}(x) Y_{\ell m}^{\pm}(\theta,\phi)
\end{equation}
where $Y_{\ell m}^{\pm}(\theta,\phi)$ are the linear combinations of spherical harmonics satisfying the boundary condition $Y_{\ell m}^{\pm}(\pi-\theta,-\phi) = \pm Y_{\ell m}^{\pm}(\theta,\phi)$.  Their explicit forms are \cite{Maru:2009wu}
\begin{eqnarray}
\label{modef1}
Y_{\ell m}^+(\theta, \phi) 
&\equiv&  \frac{(i)^{\ell+m}}{\sqrt{2}}[Y_{\ell m}(\theta, \phi) 
+ (-1)^{\ell} Y_{\ell-m}(\theta, \phi)] 
\quad \textrm{for} \quad m \not=0  \\
\label{modef2}
Y_{\ell m}^-(\theta, \phi) 
&\equiv&  \frac{(i)^{\ell+m+1}}{\sqrt{2}}[Y_{\ell m}(\theta, \phi) 
- (-1)^{\ell} Y_{\ell-m}(\theta, \phi)] 
\quad \textrm{for} \quad m \not=0  \\
\label{modef3}
Y_{\ell0}^{+(-)}(\theta) 
&\equiv&
\left\{\begin{array}{l}
Y_{\ell0}(\theta) \quad \textrm{for} \quad m=0 \ \textrm{and} \ 
\ell=\textrm{even (odd)} \\
0 \qquad \quad \textrm{for} \quad m=0 \ \textrm{and} \ 
\ell=\textrm{odd (even)}.
\end{array}\right. 
\end{eqnarray}
Note that we do not have KK mode functions that are odd under $\phi \rightarrow \phi + 2 \pi$ since the KK modes are specified by the integer angular momentum quantum numbers $\ell$ and $m$ of gauge field $A_M$ on the two-sphere.  Thus, the components of $A_{\mu}$ and $A_{\theta,\phi}$ with $(+,-)$ or $(-,+)$ parities do not have corresponding KK modes.
Applying the KK expansion and integrating about $d \Omega$, we obtain the kinetic and KK mass terms for the KK modes of $A_{\mu}$
\begin{eqnarray}
\label{masstermAm}
L_M 
&=& -\frac{1}{2} \left[
\partial_{\mu} A^{\ell m}_{\nu}(x)-\partial_{\nu}A^{\ell m}_{\mu}(x)
\right]
\left[
\partial^{\mu}A^{\ell m \nu}(x)-\partial^{\nu}A^{\ell m \mu}(x)
\right] \nonumber \\
&& \qquad
 + \frac{\ell(\ell+1)}{R^2} A_{\mu}^{\ell m}(x)  A^{\ell m \mu}(x)  \nonumber \\
&& \qquad + \frac{9 q_I^2}{R^2} \biggl[ \int d \Omega \frac{(\cos \theta \pm 1)^2}{\sin^2 \theta} (Y_{\ell m}^{\mp})^2 \biggr] A_{\mu}^{\ell m}(x) A^{\ell m \mu}(x) ~,
\end{eqnarray}
where we have used $Tr[Q_i Q^i]=2$ and $[A_{\mu}(x),Q_I] = q_I (A_{\mu}^i(x) Q_i - A_{i \mu}(x) Q^i )$.  Therefore, the KK masses of $A_\mu$ are 
\begin{eqnarray}
\label{KKmass-gauge}
M_{\ell m}^{KK}(A_\mu) &=& 
\frac{1}{R} \sqrt{\ell(\ell+1)+(m^B_{\ell m})^2} ~, \\
(m^B_{\ell m})^2 &=& 
9 q_I^2 \int d \Omega 
\frac{(\cos \theta \pm 1)^2}{\sin^2 \theta} (Y_{\ell m}^{\mp})^2 ~,
\end{eqnarray}
where $m^B_{\ell m}$ corresponds to the contribution from the background gauge field.  Note that Eq.~(\ref{KKmass-gauge}) agrees with Eq.~(\ref{eq:nonSMgaugeMass}) when $\ell = 0$.  Also, since the SM gauge bosons have $q_I = 0$, their KK masses are simply $\sqrt{\ell(\ell+1)}/R$ at the tree level.

The kinetic and KK mass terms of $A_{\theta}$ and $A_{\phi}$ are obtained from the terms in the higher dimensional gauge sector
\begin{eqnarray}
\label{scalar}
L &=& \frac{1}{2 g^2} \int d \Omega \Biggl\{ \Bigl( Tr[(\partial_{\mu} A_{\theta}-i[A_{\mu},A_{\theta}])^2] 
+ Tr[(\partial_{\mu} A_{\theta}-i[A_{\mu},\tilde{A}_{\phi}])^2 ]  \Bigr) \nonumber \\
&& \qquad \qquad -  \frac{1}{R^2} Tr \biggl[  \biggl( \frac{1}{\sin \theta} \partial_{\theta} (\sin \theta \tilde{A}_{\phi} 
+ \sin \theta \tilde{A}^B_{\phi}) \nonumber \\
&& \qquad \qquad \qquad \qquad
-\frac{1}{\sin \theta} \partial_{\phi} A_{\theta} 
- i[A_{\theta},\tilde{A}_{\phi}+\tilde{A}^B_{\phi}] \biggr)^2 \biggr] \Biggr\} ~. \nonumber \\
\end{eqnarray}
The first line on the right-hand side of Eq.~(\ref{scalar}) corresponds to the kinetic terms, and the second line corresponds to the potential term.  Applying the background gauge field Eq.~(\ref{background}), the potential becomes
\begin{eqnarray}
L_V = -\frac{1}{2 g^2 R^2} \int d \Omega Tr \biggl[ 
\biggl( \frac{1}{\sin \theta} \partial_{\theta} (\sin \theta \tilde{A}_{\phi}) + Q - \frac{1}{\sin \theta} \partial_{\phi} A_{\theta} -i [A_{\theta}, \tilde{A}_{\phi}+\tilde{A}_{\phi}^B] \biggr)^2 
\biggr]
\end{eqnarray}
For $A_{\theta}$ and $A_{\phi}$ we use the following KK expansions to obtain the KK mass terms,
\begin{eqnarray}
\label{expansion3}
A_{\theta}(x,\theta,\phi) 
&=& \sum_{\ell m (\neq 0)} \frac{-1}{\sqrt{\ell(\ell+1)}} \bigl[ \Phi_1^{\ell m}(x) \partial_{\theta} Y_{\ell m}^{\pm}(\theta,\phi) 
+ \Phi_2^{\ell m}(x) \frac{1}{\sin \theta} \partial_{\phi} Y_{\ell m}^{\pm}(\theta,\phi)  \bigr] ~, \nonumber \\ \\
\label{expansion4}
A_{\phi}(x,\theta,\phi) 
&=& \sum_{\ell m (\neq 0)}\frac{1}{\sqrt{\ell(\ell+1)}} 
\bigl[ \Phi_2^{\ell m}(x) \partial_{\theta} Y_{\ell m}^{\pm}(\theta,\phi)
  - \Phi_1^{\ell m}(x) \frac{1}{\sin \theta} \partial_{\phi} Y_{\ell m}^{\pm}(\theta,\phi)  \bigr] ~, \nonumber \\
\end{eqnarray}
where the factor of $1/\sqrt{\ell(\ell+1)}$ is needed for normalization.  These particular forms are convenient in giving diagonalized KK mass terms \cite{Maru:2009wu}.  Applying the KK expansions Eq.~(\ref{expansion3}) and Eq.~(\ref{expansion4}), we obtain the kinetic term
\begin{eqnarray}
L_{K} = \frac{1}{2 g^2} \sum_{\ell m (\neq 0)} Tr \biggl[ \partial_{\mu} \Phi_1^{\ell m}(x) \partial^{\mu} \Phi_1^{\ell m}(x) +  \partial_{\mu} \Phi_2^{\ell m}(x) \partial^{\mu} \Phi_2^{\ell m}(x) \biggr]
\end{eqnarray}
where only terms quadratic in $\partial_{\mu} \Phi$ are retained.  The potential term
\begin{eqnarray}
L_V 
&=& -\frac{1}{2 g^2 R^2} \sum_{\ell m (\neq 0)} \int d \Omega 
Tr \biggl[  \biggl( \frac{\Phi_{2}^{\ell m}}{\sqrt{\ell(\ell+1)}} 
\frac{1}{\sin \theta} \partial_{\theta} 
(\sin \theta \partial_{\theta} Y_{\ell m}^{\pm} ) +Q \nonumber \\
&& \qquad \qquad
+ \frac{\Phi_{2}^{\ell m}}{\sqrt{\ell(\ell+1)}} \frac{1}{\sin^2 \theta} \partial_{\phi}^2 Y_{\ell m}^{\pm} \nonumber \\ 
&& \qquad \qquad
- \frac{i}{\ell(\ell+1)} \Bigl[ - \Phi_{1}^{\ell m} \partial_{\theta} Y_{\ell m}^{\pm} - \Phi_{2}^{\ell m} \frac{1}{\sin \theta} \partial_{\phi} Y_{\ell m}^{\pm}, 
\nonumber \\
&& \qquad \qquad
\Phi_{2}^{\ell m} \partial_{\theta} Y_{\ell m}^{\pm} - \Phi_{1}^{\ell m} \frac{1}{\sin \theta} \partial_{\phi} Y_{\ell m}^{\pm} + \sqrt{\ell(\ell+1)} A_{\phi}^B \Bigr] \biggr)^2 \biggr] ~,
\end{eqnarray}
where only terms diagonal in $(\ell,m)$ are consider.  Using the relation $\frac{1}{\sin \theta} \partial_{\theta} (\sin \theta \partial_{\theta} Y_{\ell m}) + \frac{1}{\sin^2 \theta} \partial_{\phi}^2 Y_{\ell m} = -\ell(\ell+1)Y_{\ell m}$, the potential term is simplified as
\begin{eqnarray}
L_V &=& -\frac{1}{2 g^2 R^2} \sum_{\ell m(\neq 0)} \int d \Omega Tr \biggl[ \biggl( 
-\sqrt{\ell(\ell+1)}  \Phi_2^{\ell m}Y_{\ell m}^{\pm} +Q \nonumber \\
&& \qquad \qquad \qquad 
+\frac{i}{\ell(\ell+1)} [\Phi_1^{\ell m}, \Phi_2^{\ell m}]  \bigl( \partial_{\theta} Y_{\ell m}^{\pm} \partial_{\theta} Y_{\ell m}^{\pm}
+\frac{1}{\sin^2 \theta} \partial_{\phi} Y_{\ell m}^{\pm} \partial_{\phi} Y_{\ell m}^{\pm} \bigr) \nonumber \\
&& \qquad \qquad \qquad 
+\frac{i}{\sqrt{\ell(\ell+1)}} [\Phi_1^{\ell m}, \tilde{A}_{\phi}^B] \partial_{\theta} Y_{\ell m}^{\pm} 
+ \frac{i}{\sqrt{\ell(\ell+1)}} [\Phi_2^{\ell m}, \tilde{A}_{\phi}^B] \frac{\partial_{\phi} Y_{\ell m}^{\pm}}{\sin \theta} \biggr)^2 \biggr] ~. \nonumber \\
\end{eqnarray}
To obtain the mass term, we focus on terms quadratic in $\Phi_{1,2}$: 
\begin{eqnarray}
\label{massterm}
L_M &=& -\frac{1}{2 g^2 R^2} \int d \Omega Tr \biggl[ 
\ell(\ell+1) (\Phi_2^{\ell m})^2 (Y_{\ell m}^{\pm})^2 \nonumber \\
&& \qquad
+ \frac{2i Q}{\ell(\ell+1)} [\Phi_1^{\ell m},\Phi_2^{\ell m}] 
\Bigl( \partial_{\theta} Y_{\ell m}^{\pm} \partial_{\theta} Y_{\ell m}^{\pm} + \frac{1}{\sin^2 \theta} \partial_{\phi} Y_{\ell m}^{\pm} \partial_{\phi} Y_{\ell m}^{\pm} \Bigr) \nonumber \\
&& \qquad
+2 i \tilde{A}_{\phi}^B [\Phi_1^{\ell m},\Phi_2^{\ell m}] Y_{\ell m}^{\pm} \partial_{\theta} Y_{\ell m}^{\pm} 
- \frac{1}{\ell(\ell+1)} [\Phi_1^{\ell m},\tilde{A}_{\phi}^B]^2 (\partial_{\theta} Y_{\ell m}^{\pm})^2 \nonumber \\
&& \qquad
- \frac{1}{\ell(\ell+1)} [\Phi_2^{\ell m},\tilde{A}_{\phi}^B]^2 \frac{(\partial_{\phi} Y_{\ell m}^{\pm})^2}{\sin^2 \theta} \biggr]. \nonumber \\
\end{eqnarray}
Note that we have dropped the term proportional to $[\Phi_1,\tilde{A}_{\phi}^B] [\Phi_2,\tilde{A}_{\phi}^B]$ because this term vanishes after turning the field into the linear combinations of $\Phi$ and $\Phi^\dagger$, Eqs.~(\ref{okikae1}) and (\ref{okikae1}): 
\begin{eqnarray}
Tr[[\Phi_1,\tilde{A}_{\phi}^B][\Phi_1,\tilde{A}_{\phi}^B]] &\rightarrow& Tr[[(\Phi+\Phi^{\dagger}),Q] [(\Phi-\Phi^{\dagger}),Q] ] \nonumber \\
&\propto& Tr[(\Phi-\Phi^\dagger)(\Phi+\Phi^{\dagger})] \nonumber \\
&\propto& Tr[\Phi \Phi^{\dagger}] - Tr[\Phi^{\dagger} \Phi] =0
\end{eqnarray}
Integrating the second term of Eq.~(\ref{massterm}) by part, we obtain
\begin{eqnarray}
L_M &=& -\frac{1}{2g^2 R^2} \biggl( \ell(\ell+1) Tr[(\Phi_2^{\ell m})^2] +2i Tr[Q [\Phi_1^{\ell m},\Phi_2^{\ell m}] ] \nonumber \\
&& \qquad \qquad  -2 i Tr[Q [\Phi_1^{\ell m},\Phi_2^{\ell m}] ]  \int d \Omega \frac{\cos \theta \mp 1}{\sin \theta} Y_{\ell m}^{\pm} \partial_{\theta} Y_{\ell m}^{\pm} \nonumber \\
&& \qquad \qquad -\frac{1}{\ell(\ell+1)} [\Phi_1^{\ell m},Q]^2 \int d \Omega \frac{(\cos \theta \mp1)^2}{\sin^2 \theta} (\partial_{\theta} Y_{\ell m}^{\pm})^2 \nonumber \\
&& \qquad \qquad - \frac{1}{\ell(\ell+1)} [\Phi_2^{\ell m},Q]^2 \int d \Omega \frac{(\cos \theta \mp 1)}{\sin^2 \theta} \frac{(\partial_{\phi} Y_{\ell m}^{\pm})^2}{\sin^2 \theta} \biggr) ~. \nonumber \\
\end{eqnarray}
Therefore, the KK masses depend on the U(1)$_I$ charges of the scalar fields.

For components with zero U(1)$_I$ charge, we write $\Phi_{1(2)}(x)$ as $\phi_{1(2)}(x) Q$ where $Q$ is the corresponding generator of E$_6$ in Eq.~(\ref{d78}) with zero U(1)$_I$ charge.  Taking the trace, we have the following kinetic and KK mass terms instead:
\begin{eqnarray}
L =  \sum_{\ell m(\neq 0)} \biggl( \partial_{\mu} \phi_1^{\ell m}(x) \partial^{\mu} \phi_1^{\ell m}(x) +  \partial_{\mu} \phi_2^{\ell m}(x) \partial^{\mu} \phi_2^{\ell m}(x)  
+ \ell(\ell+1) \phi_2^{\ell m}(x) \phi_2^{\ell m}(x) \biggr)
\end{eqnarray}
where we have made the substitution $\phi_i \rightarrow g \phi_i$.  Note that $\phi_1$ is considered as a massless Nambu-Goldstone (NG) boson in this case.

For components with nonzero U(1)$_I$ charge, we use Eq.~(\ref{okikae1}) and (\ref{okikae2}) and write $\Phi(x)$ as $\phi^i(x)Q_i$ where $Q_i$ is the corresponding generator of $E_6$ in Eq.~(\ref{d78}) with nonzero U(1)$_I$ charge.  The commutator between $Q$ and $\Phi$ is
\begin{equation}
[Q,\Phi] = 3[Q_I,Q_i] \phi^i = 3 q_I \phi^i ~,
\end{equation}
where we have used $Q=3Q_I$ and that $q_I$ is a constant determined by the U(1)$_I$ charge of the corresponding component.  Finally, the Lagrangian becomes
\begin{eqnarray}
L &=&  \sum_{\ell m(\neq 0)} \biggl\{ \partial_{\mu} \phi^{\dagger}_{\ell m}  \partial^{\mu} \phi_{\ell m} \nonumber \\
&& \qquad \quad -\frac{1}{4 R^2} \biggl[ 
2 \ell(\ell+1) \phi_{\ell m}^\dagger \phi_{\ell m}
-12 q_I \phi_{\ell m}^\dagger \phi_{\ell m} \nonumber \\
&& \qquad \qquad \qquad
+12 q_I \phi_{\ell m}^\dagger \phi_{\ell m } 
\int d \Omega \frac{\cos \theta \mp 1}{\sin \theta} Y_{\ell m}^{\pm} \partial_{\theta} Y_{\ell m}^{\pm} \nonumber \\
&& \qquad \qquad \qquad 
+\frac{18 q_I^2}{ \ell(\ell+1)} \phi_{\ell m}^\dagger \phi_{\ell m} 
\int d \Omega \frac{(\cos \theta \mp1)^2}{\sin^2 \theta} 
\left(
(\partial_{\theta} Y_{\ell m}^{\pm})^2
+ \frac{(\partial_{\phi} Y_{\ell m}^{\pm})^2}{\sin^2 \theta}
\right) \biggr] \biggr\}. \nonumber \\
\end{eqnarray}
where the subscript $i$ is omitted for simplicity.  The KK masses of the complex scalar field $\phi$ are then
\begin{eqnarray}
\label{KKmass-scalar}
M_{\ell m}^{KK}(\phi) &=& \frac{1}{R} \sqrt{\frac{\ell(\ell+1)}{2}+(m_{\ell m}^B)^2} ~, \nonumber \\
(m_{\ell m}^B)^2 &=& -3 q_I  +3 q_I  \int d \Omega \frac{\cos \theta \mp 1}{\sin \theta} Y_{\ell m}^{\pm} \partial_{\theta} Y_{\ell m}^{\pm} \nonumber \\
&& \qquad
+\frac{9 q_I^2}{ 2\ell(\ell+1)} \int d \Omega \frac{(\cos \theta \mp1)^2}{\sin^2 \theta} (\partial_{\theta} Y_{\ell m}^{\pm})^2 \nonumber \\
&& \qquad + \frac{9 q_I^2}{2\ell(\ell+1)}  \int d \Omega \frac{(\cos \theta \mp 1)^2}{\sin^2 \theta} \frac{(\partial_{\phi} Y_{\ell m}^{\pm})^2}{\sin^2 \theta} ~.
\end{eqnarray}
The squared KK mass $\left( M_{\ell m}^{KK} \right)^2$ is always positive except for the lowest mode ($\ell=1,m=1$).  In fact, the squared KK mass of the $(1,1)$ mode agrees with the coefficient of quadratic term in the Higgs potential (\ref{H-potential}).


\section{Summary and discussions}
\label{summary}

We have reviewed a gauge theory defined on 6D spacetime with the $S^2/Z_2$ topology on the extra space.
Two scenarios are considered to construct a 4D theory from the 6D model. 
One scenario based on the SO(12) gauge group requires a symmetry condition for the gauge field.  The other involves the E$_6$ gauge group, but does not need the symmetry condition.
Non-trivial boundary conditions on the extra space are imposed in both scenarios.


We explicitly give the prescriptions to identify the gauge field and the scalar field remaining in 4D spacetime after the dimensional reduction.
We show that the SU(3)$_C\times$ SU(2)$_L$ $\times$ U(1)$_Y$ $\times$ U(1)$_X$ $\times$ U(1)$_I$ gauge symmetry remains in 4D spacetime, and that the SM Higgs doublet with a suitable potential for electroweak symmetry breaking can be derived from the gauge sector in both models.  The Higgs boson mass is also predicted in such models.
Massless fermion modes are also successfully obtained as the SM fermions by introducing appropriate field contents in 6D spacetime, with suitable parity assignments on the $S^2/Z_2$ extra dimension, and incorporating the background gauge field.
We also discuss about the massive KK modes of fermions for the scenario with the symmetry condition and the KK modes of all fields for the one without the symmetry condition.  The lightest fermonic KK mode can serve as a dark matter candidate.
In general, they may give rise to rich phenomena in collider experiments and implications in cosmological studies. 
%


To make our models more realistic, there are several challenges such as
eliminating the extra U(1) symmetries 
and
constructing the realistic Yukawa couplings, 
which are the same as other gauge-Higgs unification models.
We, however, can get Kaluza-Klein modes in our models.
This suggests that we obtain the dark matter candidate in our model.
Thus it is very important to study these models further such as dark matter physics and collider physics.
\section*{Acknowledgement}

This research was supported in part by 
the Grant-in-Aid for the Ministry
of Education, Culture, Sports, Science, and Technology,
Government of Japan, No. 20540251 (J.S.), and
the National Science Council of R.O.C. under Grant No. NSC-100-2628-M-008-003-MY4.

\end{document}